\documentstyle[aps,prl,twocolumn,floats,epsf]{revtex}

\begin{document} 
\draft 
\title{Random Berry Phase Magnetoresistance as a \\
Probe of Interface Roughness in Si MOSFETs} 
\author{H. Mathur$^{1}$ and Harold U. Baranger$^{2}$} 
\address{$^{1}$Department of Physics,
Case Western Reserve University, Cleveland OH 44106-7079} 
\address{$^{2}$Department of Physics,
Duke University, Box 90305, Durham NC 27708-0305}

\date{\today} 
\maketitle


\begin{abstract} 
The effect of silicon-oxide interface roughness on the weak-localization
magnetoconductance of a silicon MOSFET in a magnetic field, tilted with
respect to the interface, is studied. It is shown that an electron picks
up a random Berry's phase as it traverses a closed orbit. Effectively, due
to roughness, the electron sees an uniform field parallel to the interface
as a random perpendicular field. At zero parallel field the dependence of
the conductance on the perpendicular field has a well known form, the
weak-localization lineshape. Here the effect of applying a fixed parallel
field on the lineshape is analyzed. Many types of behavior are found
including homogeneous broadening, inhomogeneous broadening and a
remarkable regime in which the change in lineshape depends only on the
magnetic field, the two length scales that characterize the interface
roughness and fundamental constants. Good agreement is obtained with
experiments that are in the homogeneous broadening limit. The implications
for using weak-localization magnetoconductance as a probe of interface
roughness, as proposed by Wheeler and coworkers, are discussed.

\end{abstract} 
\pacs{PACS: }



\section{Introduction}

Disorder has a profound effect on electron transport at low temperature.
The scaling theory of localization applies over an enormous domain and is
a keystone in our understanding of disorder effects \cite{lee,rammer}. For
two-dimensional samples that are weakly-disordered and at low temperature
(``weak-localization regime'') the theory predicts the precise dependence
of the conductance on an applied magnetic field (``weak-localization
lineshape'') \cite{hikami}. The exquisite agreement of the predicted
lineshape with experiment constitutes an important confirmation of scaling
theory \cite{bergmann}.

Silicon MOSFETs are an important experimental realization of a
two-dimensional electronic system. In a MOSFET electrons are confined to
the interface between layers of oxide and semiconductor. In this paper we
analyze the effect of interface roughness on the weak-localization
lineshape in MOSFETs. Although the effects are small they are of interest
from various points of view:

i) The deviations from the known lineshape of an ideal interface are small
but measurable: our results agree well with the experiments that
stimulated our work \cite{bob1,bob2,albert}.

ii) The relevant effect of roughness on the electrons can be traced to a
subtle quantum interference effect: Berry's phase \cite{berry,geomphase}. There
has been considerable theoretical interest in designing an experiment
sensitive to the influence of Berry's phase on quantum electron transport
\cite{ady}. Here we have identified a prior experimental detection of it.

iii) A remarkable feature of many quantum transport phenomena is their
universality, in the sense that the observed effects are independent of
microscopic sample properties such as the mean mobility. The effects of
roughness have this feature. We find a particularly striking regime in
which the effect is determined entirely by the geometric parameters of the
interface and the applied magnetic field while being independent of both 
sample mobility and temperature (which controls the dephasing length).

iv) There has been much experimental and theoretical work on the motion of
electrons in a random magnetic field, particularly in the strong field
limit~\cite{bending,geim,smith,weiss,mancoff,gusev,theory} (motivated in
part by possible relevance to cuprates and the quantum Hall system at
filling factor 1/2). We find that due to roughness electrons see a
uniform in-plane magnetic field as a random perpendicular field; hence it
may be possible to use this system to study electron motion in a random
magnetic field. Here our analysis is restricted to the weak field limit
appropriate for calculating the lineshape in MOSFETs. Experiments on other
realizations of the random field problem are briefly discussed in Section VI.

v) Finally this work may have practical implications. MOSFETs are the
building blocks of modern electronics. Roughness of the interface between
the oxide and semiconductor influences the mobility of the device and has
been correlated with device failure due to dielectric breakdown of the
oxide\cite{MOSFET1,MOSFET2}. It is therefore of technological interest to
characterize the roughness. R. G. Wheeler and co-workers have proposed
that magnetoresistance measurements on a MOSFET in the weak-localization
regime can be used as a non-destructive probe of its interface roughness
\cite{bob1,bob2}. They and others have carried out experiments that
demonstrate the feasibility of making the needed measurements
\cite{bob1,bob2,albert}. The present work contributes to this program by
providing the precise relationship between the interface roughness
parameters and the measured magnetoresistance.

\begin{figure}[t]
\begin{center}
\leavevmode
\epsfxsize=3.0in \epsfbox{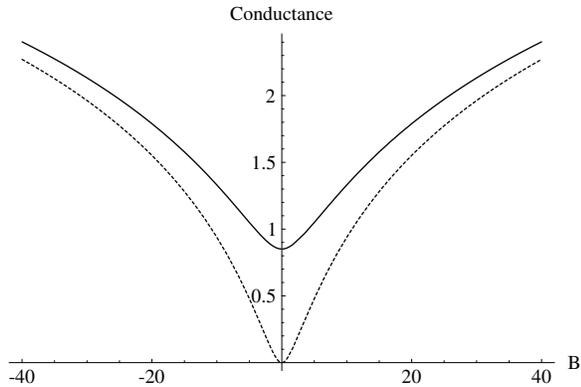}
\end{center}
\caption{MOSFET with short-ranged correlated roughness
($L \ll l_e$). The conductance (in units of $10^{-5}$ mhos) is 
plotted as a function of the applied perpendicular field, $B_{\perp}/
B_{\phi}$. The dotted curve is the classic weak-localization lineshape.
It corresponds to zero in-plane field, $B_{\parallel} = 0$. The 
solid curve shows the effect of applying a fixed in-plane magnetic
field. It corresponds to $ B_e/B_{\phi} = 400 $ and
$B_{\parallel}/B_{\phi} = ( \sqrt{8}/ \pi^{1/4} ) \sqrt{ l_e l_{\phi}^2 /
L \Delta^2 }$. }
\end{figure}

Figs. 1 and 2 summarize our findings. Following the experiments of 
Ref.~\onlinecite{bob1,bob2} we imagine the device is placed in a magnetic field
tilted with respect to the plane of the interface. The in-plane component
of the field, $B_{\parallel}$ is kept fixed and the conductance is plotted
as a function of the perpendicular component, $B_{\perp}$.  To understand
these curves it is useful to recall that there are two important length
scales that determine the transport properties of the sample: $l_{e}$, the
elastic mean free path of electrons (which determines the mobility), and
$l_{\phi}$, the distance over which electrons maintain phase-coherence and
are thus able to interfere. In the weak-localization regime, $l_{\phi} \gg
l_{e}$. Also, atomic force microscope images reveal that statistically the
rough interface can be characterized by two parameters: $\Delta=$ the
root-mean-square height fluctuations and $L=$ the distance over which the
fluctuations are correlated.

\begin{figure}[t]
\begin{center}
\leavevmode
\epsfxsize=3.0in \epsfbox{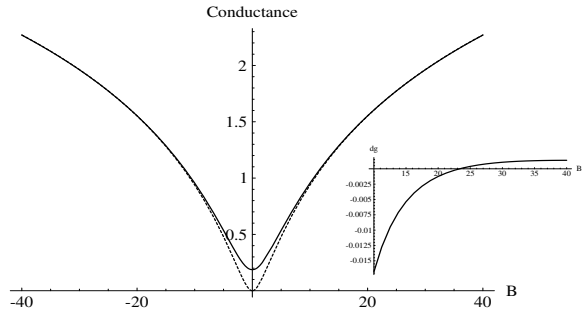}
\end{center}
\caption{Same as Fig. 1 for a MOSFET with long-range correlated
roughness ($L \gg l_e$). The dotted curve corresponds to $B_{\parallel}
= 0$; the solid curve, to $B_{\parallel} \Delta L = \phi_{{\rm sc}}/10$. 
Also $ B_e = 400 B_{\phi}$ and $L = l_{\phi}/\sqrt{5}$.
The crossing of the two curves is highlighted by a plot of their 
difference (inset).}
\end{figure}

For $B_{\parallel}=0$ we obtain the classic weak-localization lineshape
appropriate for an ideal interface (dotted curves in Figs. 1 and 2). The
height\footnote{The height is measured relative to the Drude conductance
given by the large-field asymptotic value of the conductance corrected for
classical magnetoresistance.}and the width of the peak are controlled by
the dephasing length $l_{\phi}$. The solid curves show possible lineshapes
when the parallel field is turned on. Fig. 1 shows that when the roughness
is correlated over a very short length scale ($L \ll l_{e}$) the effect of
the parallel field is to decrease the dephasing length; the lineshape is
otherwise unaltered. Borrowing the terminology of magnetic resonance and
atomic physics the effect of the in-plane field in this limit may be
described as {\it homogeneous} line-broadening. In the opposite limit when
the interface fluctuates slowly ($L \gg l_{e}$), the effects are more
dramatic.  The deviation from the ideal lineshape changes sign as a
function of perpendicular field (Fig. 2, inset). Again, by analogy to
magnetic resonance and atomic physics, the extreme limit, $ L \gg l_{\phi}
\gg l_{e} $, can be interpreted as {\it inhomogeneous} broadening (see
Section V). The sign change is then traceable to a necessary inflection
point in the ideal weak-localization lineshape. The intermediate regime $
l_{\phi} \gg L \gg l_{e} $ is particularly interesting although it cannot
be so simply interpreted. In this limit the deviation from the ideal
lineshape has a universal form independent of $ l_{e} $ and $ l_{\phi}
$.  For example, the sign change occurs at a value of the perpendicular
field determined by the purely geometric condition $B_{\perp} L^2 =
(1.79...) h/e$. Note that in all cases the deviation from the ideal
lineshape grows with $B_{\parallel}$ and $\Delta$ (in proportion to
$B_{\parallel}^2$ and $\Delta^2$ in the experimentally relevant regime).

Further discussion and a detailed summary 
of our results is given in Section VII.
 
\section{Born-Oppenheimer Analysis}

As a model of the silicon-oxide interface, for many purposes it is
sufficient to regard the oxide as an impenetrable hard wall and to assume
that the electrons are bound to the interface by a uniform perpendicular
electric field \cite{ando}. It will become apparent in the sequel that for
the present purpose it is only necessary to assume that the electrons are
firmly bound to the interface: it is not necessary to commit to a
specific form of the confinement potential $U_{{\rm conf}}$. If the axes
are chosen so that the interface lies in the plane $ z=0 $, for an ideal
interface the confinement potential would depend only on $z$; but for a
rough interface it would vary from point to point and hence also depend on
$x$ and $y$.  The strength of the confining potential controls the extent
of the wavefunction in the $z$ direction, denoted $\ell$.  We also allow for
the possibility of electron scattering by impurities in the semiconductor
by introducing a potential $U_{{\rm imp}}$.  The magnetic field is assumed
to lie in the $x$-$z$ plane: thus $ B_{x} = B_{\parallel}, B_{y} = 0$ and
$B_{z} = B_{\perp}$. It is convenient to work in a Landau gauge and to
choose $A_{x} = 0, A_{y} = B_{\perp} x$ and $ A_{z} = B_{\parallel} y$.
The Schr\"{o}dinger equation is
\begin{eqnarray}
   - \frac{\hbar^2}{2m} \left( \frac{\partial}{\partial z} - 
   i e \frac{A_{z}}{\hbar} \right)^2 \Psi
   - \frac{\hbar^2}{2m} \left( \nabla -
   i e \frac{{\mathbf A}}{\hbar} 
   \right)^2 \Psi & & \nonumber \\
   + \,
   U_{{\rm conf}}(z, x, y) \Psi +
   U_{{\rm imp}}(z, x,y) \Psi = E \Psi(x,y,z). & &
\end{eqnarray}
For later convenience $\nabla$ is used to denote the two-dimensional
gradient in the $x$-$y$ plane.  Similarly ${\mathbf A}$ denotes the $x$-$y$
component of the vector potential.

For an ideal interface the electronic motion in the plane of the interface
and in the perpendicular direction separate. The motion in the
perpendicular $z$-direction is quantized by the confining potential.
Provided the temperature is sufficiently low and the density of electrons
not too high, only the lowest subband mode in the $z$-direction is
populated and the electronic motion is essentially two-dimensional. For a
rough interface the motion no longer separates but we find it is an
excellent approximation to integrate out the motion perpendicular to the
interface and to obtain an effective Hamiltonian for motion in the plane
of the interface using the Born-Oppenheimer method \cite{shankar}.

For simplicity first suppose there is no magnetic field.  Assume that the
motion in the $x$-$y$ plane is slow; the motion perpendicular to it, fast.
The first step in the Born-Oppenheimer method is to analyze the motion of
the fast co-ordinates treating the slow co-ordinates as parameters. In
this case it is necessary to solve
\begin{eqnarray}
   - \frac{\hbar^2}{2m} \frac{d^2}{d z^2} \phi_{n}(z; x, y) 
   + U_{{\rm conf}} (z, x, y) \phi_{n}(z;x,y) & & \nonumber \\
   =
   E_{n}(x,y) \phi_{n} (z; x, y) & &
\end{eqnarray}
Here $\phi_{n}(z;x,y)$ denotes the ``local subband'' wavefunction and
$E_{n}(x,y)$ is the energy of the $n^{{\rm th}}$ subband wavefunction.
Sometimes it will be convenient to write the sub-band states using Dirac
notation: $ \phi_{n}(z;x,y) \rightarrow | n; x,y \rangle$.  Both the
subband wavefunction and the subband energy vary with location in the
$x$-$y$ plane because the confinement potential varies from point to
point. It is assumed that the subband wavefunction is normalized
everywhere so that $\langle n;x,y|n;x,y \rangle = \int_{-\infty}^{\infty}
d z |\phi_{n} (z;x,y) |^2 = 1$.  For a given $(x,y)$ Eq. (2) only fixes
the subband wavefunction up to a phase. It is convenient to choose the
wavefunctions to be real and to vary smoothly with $x$ and $y$.

According to the Born-Oppenheimer method an approximate solution
to Eq. (1) is
\begin{equation}
   \Psi(x,y,z) \approx \psi(x,y) \phi_{g}(z;x,y).
\end{equation}
Here $\phi_{g}$ denotes the lowest energy subband wavefunction
and $\psi(x,y)$ is governed by
\begin{eqnarray}
   - \frac{\hbar^2}{2m} \nabla^2 \psi(x,y) 
   + \tilde{U}_{{\rm imp}} (x,y) \psi(x,y)
   & & \nonumber \\
   + E_{g} (x,y) \psi (x,y) + W_{g} (x,y) \psi(x,y) & = & E \psi ( x, y ).
\end{eqnarray}
Eq. (4) describes the motion of the electrons in the plane of the interface
after the transverse motion has been integrated out.  $\tilde{U}_{\rm
imp}$, the effect of impurities, now depends only on $x$ and $y$. For
impurities that are sufficiently far from the interface so that their
potential does not vary significantly over the confinement scale $ l $,
$\tilde{U}_{\rm imp} = U_{\rm imp}$, while for short-range impurities
$\tilde{U}_{\rm imp}$ incorporates their effect on electrons in the lowest
subband.  Note that the local subband energy, $E_{g}(x,y)$ appears as a
potential in the Born-Oppenheimer effective Hamiltonian.  The effective
potential $W_{g} (x,y)$ is also determined by the solution to the fast
problem. It is given by a more complicated expression 
\begin{equation}
   W_{g} (x,y) = \sum_{n \neq g} 
   \langle g;x,y| \bigl( \nabla | n;x,y \rangle \bigr)
   \cdot \langle n; x,y | \nabla | g; x,y \rangle.  
\end{equation}

Next suppose that the magnetic field is turned on. The fast co-ordinate
is now governed by
\begin{eqnarray}
& & - \frac{\hbar^2}{2m} \left( \frac{d}{dz} - i \frac{e}{\hbar} B_{\parallel}
y \right)^2 \phi_{n}(z;x,y) \nonumber \\
& & + U_{{\rm conf}}(z,x,y) \phi_{n}(z;x,y)
= E_{n}(x,y) \phi_{n}(z;x,y). 
\end{eqnarray}
Let $ \xi_{n} (z;x,y) $ denote a normalized solution to
Eq. (6) when the magnetic field is turned off. This solution is chosen
to be real and to vary smoothly with $x$ and $y$. It is easy to verify
that a solution to Eq. (6) with the magnetic field is
\begin{equation}
\phi_{n}(z;x,y) = \exp \left( i \frac{e}{\hbar} B_{\parallel} y z \right)
\xi_{n} (z;x,y).
\end{equation}
Turning on the magnetic field leaves the subband energy, $E_{n}(x,y)$
unchanged; but it introduces a non-trivial twist in the subband
wavefunctions due to the phase factor in Eq. (7).  Following Berry
\cite{berry,geomphase}, this can be made explicit by defining the
geometric vector potential $ {\mathbf A}_{g}(x,y) \equiv i (\hbar/e)
\langle g;x,y| \nabla | g;x,y \rangle $ and its curl, the geometric
magnetic field.  Straightforward calculation reveals that the cartesian
components of ${\mathbf A}$ are
\begin{eqnarray}
   A_{gx} (x,y) & = & 0; \nonumber \\
   A_{gy} (x,y) & = & - B_{\parallel} {\cal Z}_{g}(x,y)
\end{eqnarray}
and the geometric magnetic field is
\begin{eqnarray}
   B_{g}(x,y) & \equiv & \frac{\partial}{\partial x}
   A_{gy} - \frac{\partial}{\partial y} A_{gx}
   \nonumber \\
    & = & - B_{\parallel} \frac{\partial}{\partial x} {\cal Z}_{g}(x,y).
\end{eqnarray}
Here ${\cal Z}_{g}(x,y)$ is the mean $z$-coordinate for the lowest 
subband wavefunction,
\begin{equation}
   {\cal Z}_{g} (x,y) \equiv 
   \int_{-\infty}^{\infty} \!\! d z \, z \, | \xi_{g} (z;x,y) |^2.
\end{equation}

Now, according to the Born-Oppenheimer method, an approximate solution
to Eq. (1) is given by Eq. (3). Here $\phi_{g}$ denotes the lowest
energy subband wavefunction and $\psi(x,y)$ is governed by
\begin{eqnarray}
& &
- \frac{\hbar^2}{2m} \left( \nabla - i \frac{e}{\hbar} \left[ {\mathbf A}(x,y)
+ {\mathbf A}_{g}(x,y) \right] \right)^2 \psi (x,y)
\nonumber \\
& &
+ E_{g} (x,y) \psi (x,y) + W_{g}(x,y) \psi (x,y) \nonumber \\
& &
+ \tilde{U}_{{\rm imp}}(x,y) \psi(x,y) = E \psi (x,y).
\end{eqnarray}
Eq. (11) is the central result of this section. It reveals that after the
fast motion perpendicular to the interface is integrated out, the
electrons essentially move in two dimensions under the influence of an
effective potential, $\tilde{U}_{{\rm imp}} + E_{g} + W_{g}$ and an effective
perpendicular magnetic field, $B_{\perp} + B_{g}$.  

The effective magnetic
field is seen to be the component of the applied field that is
perpendicular to the surface defined by the mean $z$-coordinate of the
subband wavefunctions, ${\cal Z}_{g}(x,y)$ provided that the gradients in
this surface are small. Note that the geometric magnetic field, $B_{g}$
given by Eq. (9) is proportional to the in-plane component of the applied
field and to the gradient of the surface ${\cal Z}_{g}$. It vanishes if
$B_{\parallel} = 0$ or for an ideal interface for which the surface ${\cal
Z}_{g}$ would be flat. The effective potentials $E_{g}$ and $W_{g}$ also
owe their existence to the roughness of the interface but are independent
of $B_{\parallel}$.

Note that Eq. (6) only defines the subband wavefunctions, $\phi_{n}$ up to
a phase. A specific choice given by Eq. (7) is made in the calculation
above. This amounts to choosing a gauge for the geometric vector
potential:  a different choice of phase would transform the geometric
vector potential, ${\mathbf A}_{g}$, but would leave the geometric magnetic
field, $B_{g}$ unchanged.  Thus the calculation above uses a specific
gauge for both the applied and geometric magnetic field---the Landau gauge
defined before Eq. (1) and the gauge given by Eq. (8), respectively. The
gauges are chosen for their convenience but the results are, of course,
independent of the choice of gauge.  A fuller discussion of the gauge
invariance of the Born-Oppenheimer method is given in chapter 3.7 of 
Ref.~\onlinecite{geomphase}.

Finally we briefly discuss the applicability of the Born-Oppenheimer
approximation. In Appendix A it is shown that the approximation should
work provided $ \ell \ll \lambda_{f}$, $\nabla {\cal Z}_{g} \ll 1$ and $ \ell
\nabla^2 {\cal Z}_{g} \ll 1$. Here $\ell$ is the typical extent of the
subband wavefunction in the $z$-direction and $\lambda_{f}$ is the Fermi
wavelength of the electrons. The first condition is to ensure that the
electron density is not so high that more than one subband is occupied.
The other two inequalities are adiabaticity requirements on the surface
roughness needed to ensure that the motion in the plane of the interface
can be approximately decoupled from the transverse motion. They stipulate
that the roughness must vary slowly on a scale determined by the strength
of the confining potential.

To check the validity of the adiabaticity conditions we can use the atomic
force microscope images of the silicon-oxide interface presented in 
Ref.~\onlinecite{bob2}. These images reveal that the height fluctuations of the
interface follow a Gaussian distribution parameterized by $\Delta_{{\rm
AFM}}$, the root mean square height fluctuation, and $L$, the correlation
distance. The discussion of Appendix A suggests that the surface ${\cal
Z}_{g}$ should be similarly distributed with essentially the same
parameters. Hence the adiabaticity conditions can be expressed in terms of
the observable parameters as $\Delta_{{\rm AFM}}/L \ll 1$ and $ (\ell
\Delta_{{\rm AFM}})/L^2 \ll 1 $ and are seen to be very well satisfied for
both samples of Ref.~\onlinecite{bob2}.

\section{Weak-localization lineshape}

The purpose of this section is to review the calculation of the weak
localization lineshape. The calculation is reformulated in a way that is
suitable for the eventual goal of calculating the tilted field
magnetoresistance of a MOSFET with a rough interface.

In the previous section it was shown that effectively the electrons move
in two dimensions under the influence of a random potential, $V(x,y)$, and
a perpendicular magnetic field.  The magnetic field is the sum of the
applied perpendicular field and the geometric field, which is due to
interface roughness and the applied parallel field.
It is assumed that the surface ${\cal Z}_{g}(x,y)$ is a Gaussian
random surface with zero mean and variance given by
\begin{equation}
   \langle {\cal Z}_{g} ({\mathbf r}) {\cal Z}_{g} ({\mathbf r}') 
   \rangle_{{\rm rough}} = \Delta^2 \exp \left( - \frac{ | {\mathbf r} -
   {\mathbf r}' |^2 }{L^2}
   \right).
\end{equation}
As noted above, atomic force microscope images of the silicon-oxide
interface presented in Ref.~\onlinecite{bob2} reveal a Gaussian random
surface, which makes it extremely plausible that the surface ${\cal
Z}_{g}(x,y)$ must also be Gaussian.  An argument to this effect is given
in Appendix A. Eqs. (9) and (12) thus determine the statistics of the
random magnetic field, $B_{g}$.

The random potential $V$ is caused by impurities and by the roughness of
the interface (in the notation of the previous section $V=U_{{\rm imp}} +
E_{g} + W_{g}$). Here we shall assume that the random potential is
Gaussian white noise with zero mean and variance $\langle V({\mathbf r})
V({\mathbf r'}) \rangle_{{\rm imp}} = V_0^2 \delta^{(2)}( {\mathbf r} -
{\mathbf r'} )$. Since we assume that the interface roughness is correlated
over a length scale $L$, strictly the assumption that the random potential
is Gaussian white noise is justified only under special circumstances (if
the impurity potential is white noise and it dominates interface roughness
scattering, or if $L$ is much smaller than all relevant length scales and
the impurity scattering is either also white noise or is dominated by
interface roughness scattering), but in practice it is reasonable to
believe that our results will be more broadly applicable since
localization effects are believed to be insensitive to microscopic details
of the random potential. We estimate that for both samples studied in
Ref.~\onlinecite{bob2} impurity scattering dominates interface roughness
scattering, and for one sample $L$ is shorter than all relevant length
scales in addition; but it is expected that the results should apply under
less favorable circumstances also.

It is useful to consider the electron Green function which
obeys the Schr\"{o}dinger equation
\begin{eqnarray}
   & & \left[ - \frac{\hbar^2}{2m} 
            \left( {\mathbf \nabla} - i \frac{e}{\hbar} [ {\mathbf A} + 
   {\mathbf A}_{g} ] \right)^2 + V(x,y) - E \right] 
   {\cal G} ({\mathbf r},{\mathbf r}';E)
   \nonumber \\
   & & = - \delta^{(2)}({\mathbf r}-{\mathbf r}').
\end{eqnarray}
For the retarded Green function ${\cal G}^{R}$, the energy $E$ has a
positive infinitesimal imaginary part; for the advanced, ${\cal G}^{A}$,
a negative part. Linear response theory allows us to express the
conductance in terms of the Green functions \cite{fetter}
\begin{equation}
   g = - \frac{e^2 \hbar^3}{8 \pi m^2 L^2} 
   \int \!\!d{\mathbf r} \int \!\!d{\mathbf r}'
   \Delta {\cal G}( {\mathbf r}, {\mathbf r}', E_{f} )
   \stackrel{\leftrightarrow}{\frac{\partial}{\partial y}}
   \stackrel{\leftrightarrow}{\frac{\partial}{\partial y'}}
   \Delta {\cal G} ({\mathbf r}', {\mathbf r}, E_{f} ).
\end{equation}
Here $\Delta {\cal G} \equiv {\cal G}^{R} - {\cal G}^{A}$,
$E_{F}=$ Fermi energy and $m=$ effective mass of electrons.
We have assumed that the sample is rectangular (dimensions $L \times W$)
and current is driven along the length of the sample which is oriented
along the $y$-axis. The two-sided derivative in Eq. (14) is defined as
\begin{equation}
   f \stackrel{\leftrightarrow}{\frac{\partial}{\partial y}} g \equiv
   f \frac{\partial}{\partial y} g - g \frac{\partial}{\partial y} f.
\end{equation}
We are interested in the disorder averaged conductance. To this
end it is necessary to average products of Green functions
over the random potential and the interface roughness
under the statistical assumptions made above.

In the two subsections below we review the calculation of the
conductance at zero magnetic field and in a uniform perpendicular
field, circumstances under which there is no random magnetic
field and it is only necessary to perform an average over the
random potential. The effect of the random magnetic field is analyzed
in the following sections.

\subsection{Zero Field}

The average over the random potential 
can be calculated perturbatively \cite{rammer,abrikosov} in an expansion
in the small parameter $(E_{f} \tau_{e})^{-1}$. Here $\tau_{e}$,
the elastic scattering time for electrons, is given by $\hbar/
2 \pi \rho(E_f) V_0^2$.
where $\rho(E_{f})=$ density of states for spinless electrons.
In this approximation the average Green function is given
by \cite{abrikosov}
\begin{eqnarray}
   G^{R}({\mathbf r},{\mathbf r}', E_{f}) 
   & = & \int \frac{ d {\mathbf k} }{ (2 \pi)^2 } 
   G^{R} ( {\mathbf k}, E_{f} ) 
   \exp i {\mathbf k}.( {\mathbf r} - {\mathbf r}');
   \nonumber \\
   G^{R}( {\mathbf k}, E_{f} ) 
   & = & \left( E_{f} - \frac{ \hbar^2 k^2 }{2m} + i \frac{\hbar}{2 \tau_{e}}
   \right)^{-1}
\end{eqnarray}
Here $G({\mathbf r},{\mathbf r}';E_f) = 
\langle {\cal G}({\mathbf r},{\mathbf r}'; E_f) \rangle_{{\rm imp}}$ 
is the Green function averaged over the random potential.

The average of a product of Green functions does not factorize into a
product of the averages; it is correlated by the underlying random
potential. The correlation between ${\cal G}^{R}$ and ${\cal G}^{A}$
responsible for weak localization is called the Cooperon,
$C({\mathbf r},{\mathbf r}')$. Semiclassically this correlation can be
understood to arise from the constructive interference of classical paths
related by time reversal symmetry \cite{lee,rammer,hikami,bergmann}. 
Fig. 3 shows the Feynman diagrams that contribute to the Cooperon
\cite{lee,rammer,hikami,bergmann}. From these diagrams we see that it obeys the
integral equation
\begin{equation}
   C({\mathbf r},{\mathbf r}') = C^{(0)}({\mathbf r},{\mathbf r}') +
   \int d {\mathbf r}'' C^{(0)}({\mathbf r},{\mathbf r}'') 
      C({\mathbf r}'', {\mathbf r}').
\end{equation}
Here 
$C^{(0)}({\mathbf r},{\mathbf r}') = ( \hbar/2 \pi \rho(E_{f}) \tau_{e})
| \langle G^{R}({\mathbf r},{\mathbf r}', E_{f}) \rangle_{{\rm imp}} |^{2}$.
Using Eqs. (14) and (16), the weak-localization contribution to the
conductance, $g_{\rm WL}$, can be expressed in terms of the full 
Cooperon:
\begin{equation}
   g_{{\rm WL}} = - \frac{2}{\pi} \frac{e^2}{L^2} D \tau_{e} 
   \int \! d {\mathbf r} \, C({\mathbf r},{\mathbf r}).
\end{equation}
Here $D \equiv v_{f}^2 \tau_{e}/2$ is the
electron diffusion constant. Calculation of $g_{{\rm WL}}$ 
therefore reduces to solution of Eq. (17).

\begin{figure}[t]
\begin{center}
\leavevmode
\epsfxsize=3.0in \epsfbox{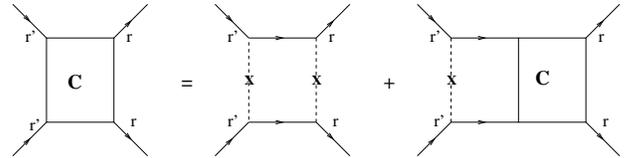}
\end{center}
\caption{Cooperon diagrams. Solid lines denote impurity 
averaged Green functions. The upper lines are retarded Green
functions; the lower lines, advanced. Dotted lines represent 
scattering from impurities (shown as crosses). }
\end{figure}

The customary procedure is to obtain eigenfunctions of $C^{(0)}$
which obey
\begin{equation}
   \int d {\mathbf r}' 
   C^{(0)}({\mathbf r},{\mathbf r}') Q_{\lambda}({\mathbf r}') = 
   \lambda Q_{\lambda} ({\mathbf r}).
\end{equation}
The solution to Eq. (17) is then 
\begin{equation}
   C({\mathbf r},{\mathbf r}')
   = \sum_{\lambda} \left( \frac{\lambda}{1-\lambda} \right) 
   Q_{\lambda}({\mathbf r}) Q_{\lambda}({\mathbf r}').
\end{equation}
It turns out that eigenfunctions that vary slowly on the scale of
$l_{e}=v_{f} \tau_{e}$ dominate the conductance; hence, it is only
necessary to accurately calculate the slowly varying solutions of Eq.
(19). Asymptotic evaluation of the Green function, Eq. (16), shows that
\begin{equation}
   C^{(0)}({\mathbf r},{\mathbf r}') \approx \frac{1}{2 \pi l_{e}^2} 
   \exp - \frac{| {\mathbf r} - {\mathbf r}' |}{l_{e}}
\end{equation}
provided $|{\mathbf r} - {\mathbf r}'| \gg \lambda_{f}$ and $k_{f} l_{e} \gg
1$. $C^{(0)}$ is therefore sharply peaked about ${\mathbf r} \approx
{\mathbf r}'$ and for the slowly varying eigenfunctions the integral in Eq.
(19) can be performed approximately by expanding $Q_{\lambda}
({\mathbf r}') $ in a Taylor's series about ${\mathbf r}' = {\mathbf r}$.
Keeping terms to second order transforms the integral equation (19) into a
diffusion equation
\begin{equation}
   D \tau_{e} \nabla^{2} Q_{\lambda}({\mathbf r})
   + Q_{\lambda}({\mathbf r}) = \lambda Q_{\lambda}({\mathbf r}).
\end{equation}
It is sufficient to use the solutions of the diffusion
Eq. (22) to construct the Cooperon, Eq. (20), and to 
calculate the conductance using Eq. (18). This completes
the usual calculation at zero field.

We now introduce an alternative formulation more suitable
for analyzing the effects of roughness. Consider the kernel
\begin{equation}
   K^{(0)}({\mathbf r},{\mathbf r}') \equiv
   \frac{1}{2 \pi l_{e}^{2}}
   \exp \left( - \frac{ |{\mathbf r} - {\mathbf r}'|^2}{2 l_{e}^{2}} \right).
\end{equation}
By analogy with $C^{(0)}$ and $C$ introduce $K$ given by 
\begin{equation}
   K({\mathbf r},{\mathbf r}') = K^{(0)}({\mathbf r},{\mathbf r}') +
   \int d {\mathbf r}'' 
   K({\mathbf r},{\mathbf r}'') K^{(0)}({\mathbf r}'',{\mathbf r}').
\end{equation}
Note that $K^{(0)}$ is not equal to $C^{(0)}$ [compare Eqs. (21) and
(23)]; but the slowly varying eigenfunctions of $K^{(0)}$ also obey the
diffusion Eq. (22). For this reason $K$ and $C$ have the same long
distance behavior. {\it Thus, we may use $K$ instead of $C$ in Eq. (18) to
calculate the weak-localization conductance since it is dominated by the
long-distance behavior.}

The chief virtue of $K$ compared to $C$ is that it can be expressed in
terms of tractable Gaussian integrals. A useful expression for $K$ is
obtained from Eq. (24) by iteration
\begin{eqnarray}
   K({\mathbf r},{\mathbf r}') & = & \sum_{n=0}^{\infty} 
   K^{(n)}({\mathbf r},{\mathbf r}') ; \nonumber \\
   K^{(n)} ({\mathbf r},{\mathbf r}') & = & 
   \int d {\mathbf r}_1 \ldots d {\mathbf r}_n
   K^{(0)}({\mathbf r},{\mathbf r}_1) K^{(0)}({\mathbf r}_1,{\mathbf r}_2) \ldots
   \nonumber \\
   & &
   \ldots K^{(0)}({\mathbf r}_n,{\mathbf r}').
\end{eqnarray}
Evidently $K^{(n)}({\mathbf r},{\mathbf r}')$ may be interpreted as the 
probability for a Gaussian random walker to go from ${\mathbf r}'$
to ${\mathbf r}$ in $n + 1$ steps.

An important physical ingredient must now be added.
As noted above, weak-localization is a quantum interference
effect. It is therefore damped by electron-electron and
electron-phonon scattering \cite{boris}. In the conventional analysis
the damping effect of these processes is included phenomenologically
by adding $\tau_{e}/\tau_{\phi}$ to the eigenvalue, $\lambda$, in
Eqs. (19), (20) and (22). Here $\tau_{\phi}$ is the dephasing time
and it is assumed that $\tau_{\phi} \gg \tau_{e}$ in the weak-localization
regime. The equivalent procedure in our formulation is to replace
\begin{equation}
   K^{(n)} \rightarrow K^{(n)} \exp \left( - \frac{(n+1) \tau_{e}}{\tau_{\phi}}
   \right).
\end{equation}
Eqs. (18), (25) and (26) then yield
\begin{equation}
   g_{{\rm WL}} = - \frac{e^2}{\hbar} \frac{D \tau_{e}}{L^2}
   \int d {\mathbf r} \sum_{n} K^{(n)}({\mathbf r},{\mathbf r}) \exp \left( - 
   \frac{n \tau_{e}}{\tau_{\phi}} \right).
\end{equation}
In summary, Eqs. (23), (25) and (27) provide the tools to calculate the
weak-localization conductance.
 
Eq. (27) has an appealing form. $g_{{\rm WL}}$ is expressed as a sum over
closed random walks, consistent with the physical idea that it is due to
the interference of closed paths and their time-reversed counterparts. The
integral over $K^{(n)}$ gives the weight of $n$-bounce paths; the
exponential factor shows that the contribution of very long paths is cut
off by dephasing.  For a different formulation of weak localization in
terms of random walks see Ref.~\onlinecite{chak_schmid}.

Let us explicitly calculate $g_{{\rm WL}}$. Successive integration
over intermediate points shows
\begin{equation}
   K^{(n)} ({\mathbf r},{\mathbf r}') = \frac{1}{(n+1) l_{e}^{2}} \exp
   \biggl( - \frac{ |{\mathbf r} - {\mathbf r}'|^2 }{(n+1) l_{e}^{2}} \biggr),
\end{equation}
a well-known result. $K^{(n)}$ has the same form as $K^{(0)}$ but with
$l_{e} \rightarrow \sqrt{n+1} \, l_e$ in agreement with the general 
proposition that the displacement of a random walker grows as the square 
root of the number of steps. Now
\begin{eqnarray}
   g_{{\rm WL}} & = & - \frac{e^2}{\hbar} \frac{W}{L}
   \sum_{n=1}^{\infty} \frac{1}{n} 
   \exp \left( - \frac{n \tau_e}{\tau_{\phi}} \right) \nonumber \\
    & \approx & - \frac{e^{2}}{\hbar} \frac{W}{L}
   \ln \frac{\tau_{\phi}}{\tau_{e}} \nonumber \\
    & = & - \frac{e^2}{\hbar} \frac{W}{L} 
   \ln \frac{l_{\phi}}{l_e},
\end{eqnarray}
a celebrated result \cite{lee,rammer,hikami}. Here we have introduced 
$l_{\phi} = \sqrt{D \tau_{\phi} }$, the dephasing length, 
and made use of the inequality $
\tau_{\phi} \gg \tau_e$. The logarithmic divergence of $g_{{\rm WL}}$ as $
l_{\phi} \rightarrow \infty$ can be traced to the slow $1/n$ decay of the
weight of long paths. In a field theory formulation it indicates that the
theory flows to strong coupling on long length scales but on shorter
scales it is ``asymptotically free'' \cite{mckane}. 
Provided that the cutoff $l_{\phi}$
is shorter than the localization length, perturbation theory is accurate.

\subsection{Uniform Perpendicular Field}

We consider magnetic fields that are weak compared to impurity scattering.
The precise condition we assume is $ B \ll B_{e}$, where $B_e = \phi_{{\rm
sc}}/2 \pi l_{e}^2$ and $\phi_{{\rm sc}} = h/2e$ is the superconducting
flux quantum. In this limit the classical paths are not modified by the
magnetic field. Its main effect is to change the phase of the paths (the
Aharonov-Bohm effect). This has no effect on the classical Drude
conductance, but the weak-localization correction, which is an
interference effect, is suppressed in a manner analyzed in
Ref.~\onlinecite{hikami}.

The starting point for this analysis is to assume that the effect of
turning on a perpendicular field on the Green
function is to multiply it by a phase factor:
\begin{equation}
G_{B_{\perp}}^{R} ({\mathbf r},{\mathbf r}';E_f)
\approx
G_{B=0}^{R} ({\mathbf r}, {\mathbf r}'; E_f)
\exp \biggl( i \frac{e}{\hbar} \!\int_{{\mathbf r}'}^{{\mathbf r}} 
d {\mathbf l} \cdot {\mathbf A} \biggr).
\end{equation}
Here $G_{B=0} $ is the disorder averaged
Green function at zero field,
Eq. (16); $G_{B_{\perp}}$, the Green function with the 
perpendicular magnetic
field turned on; and the integral in the phase factor is evaluated
along a straight line from ${\mathbf r}'$ to ${\mathbf r}$. The zeroth
order Cooperon becomes
\begin{equation}
C_{B_{\perp}}^{(0)}({\mathbf r},{\mathbf r}') =
C_{B=0}^{(0)} ({\mathbf r},{\mathbf r}')
\exp \biggl( i \frac{2e}{\hbar} \int_{{\mathbf r}'}^{{\mathbf r}} d
{\mathbf l} \cdot {\mathbf A} \biggr).
\end{equation}

Once again we are interested in the eigenfunctions of 
$C_{B_{\perp}}^{(0)}$ which obey
\begin{equation}
\int d {\mathbf r}' 
C_{B=0}^{(0)} ({\mathbf r},{\mathbf r}') 
\exp \biggl( i \frac{2e}{\hbar} \int_{{\mathbf r}'}^{{\mathbf r}}
d {\mathbf l} \cdot {\mathbf A} \biggr)
Q_{\lambda}({\mathbf r}') = \lambda Q_{\lambda}({\mathbf r}).
\end{equation}
For the slowly varying eigenfunctions Eq. (32) may be 
transformed into a differential equation by expanding
both the phase factor and $Q_{\lambda}({\mathbf r}')$ about
$ {\mathbf r}' = {\mathbf r} $ to second order. The result is 
\begin{equation}
\biggl[ D \tau_{e} \left( \nabla - i \frac{2 e}{\hbar} {\mathbf A}
({\mathbf r}) \right)^2 + 1 \biggr] Q_{\lambda} ({\mathbf r}) = 
\left( \lambda + \frac{\tau_{e}}{\tau_{\phi}} \right)
Q_{\lambda} ({\mathbf r}).
\end{equation}
In Eq. (33) we have explicitly inserted a damping term
$\tau_{e}/\tau_{\phi}$. The customary solution \cite{hikami}
is to construct the Cooperon from the Landau-level solutions to Eq. (33).

As in the previous subsection we now describe an alternative
formulation. Consider the kernel
\begin{equation}
K_{B_{\perp}}^{(0)} ({\mathbf r},{\mathbf r}') = 
K_{B=0}^{(0)} ({\mathbf r},{\mathbf r}')
\exp \left( i \frac{2e}{\hbar} \int_{{\mathbf r}'}^{{\mathbf r}} d {\mathbf l}.{\mathbf A}
\right)
\end{equation}
with $K_{B=0}^{(0)}$ given by Eq. (23). The slowly varying 
eigenfunctions of $K_{B_{\perp}}^{(0)}$ also obey Eq. (33); hence if
we define $K_{B_{\perp}}$ via
\begin{equation}
K_{B_{\perp}}({\mathbf r},{\mathbf r}') = 
K_{B_{\perp}}^{(0)}({\mathbf r},{\mathbf r}') + \int d {\mathbf r}''
K_{B_{\perp}}^{(0)} ({\mathbf r},{\mathbf r}'') 
K_{B_{\perp}} ({\mathbf r}'',{\mathbf r}'),
\end{equation}
then $K_{B_{\perp}}$ has the same long-distance behavior as 
$C_{B_{\perp}}$ and may be used in its place to calculate the 
conductance.

A useful expression for $K_{B_{\perp}}$ is obtained from Eq. (35)
by iteration
\begin{eqnarray}
K_{B_{\perp}} ({\mathbf r},{\mathbf r}') & = & \sum_{n=0}^{\infty}
K_{B_{\perp}}^{(n)} ({\mathbf r},{\mathbf r}') ;
\nonumber \\
K_{B_{\perp}}^{(n)} ({\mathbf r},{\mathbf r}') & = &
\int d {\mathbf r}_1 \ldots d {\mathbf r}_n K_{B=0}^{(0)} ({\mathbf r},{\mathbf r}_1)
\ldots \nonumber \\
& &
\ldots K_{B=0}^{(0)} ({\mathbf r}_n,{\mathbf r}') 
\exp \left( i \frac{2e}{\hbar} \int_{{\mathbf r}'}^{{\mathbf r}} 
d {\mathbf l} \cdot {\mathbf A}
\right).
\end{eqnarray}
The integral in the phase factor is evaluated along the path from
${\mathbf r}'$ to ${\mathbf r}$ obtained by joining the points ${\mathbf r}',
{\mathbf r}_n, \ldots, {\mathbf r}_1,{\mathbf r}$ with line segments in the given
order. For ${\mathbf r} = {\mathbf r}'$
\begin{eqnarray}
K_{B_{\perp}}^{(n)} ({\mathbf r},{\mathbf r}) & = & \int
d {\mathbf r}_1 \ldots d {\mathbf r}_n K_{B=0}^{(0)} ({\mathbf r},{\mathbf r}_1)
\ldots \nonumber \\
& & 
\ldots K_{B=0}^{(0)} ({\mathbf r}_n,{\mathbf r}) 
\exp \left( i 2 \pi \frac{\phi_n}{\phi_{{\rm sc}}} \right).
\end{eqnarray}
Here $ \phi_n=$ flux through the polygon defined by the closed
path from ${\mathbf r}$ to ${\mathbf r}$ and $\phi_{{\rm sc}} = h/2e$ is
the superconducting flux quantum. 

Using Eq. (18) and (36) 
\begin{equation}
g_{{\rm WL}} = - \frac{e^2}{\hbar} \frac{D \tau_{e}}{L^2}
\int \!d {\mathbf r} \sum_{n=0}^{\infty} K_{B_{\perp}}^{(n)}
({\mathbf r},{\mathbf r}) \exp \left( - (n+1) \frac{\tau_{e}}{\tau_{\phi}}
\right).
\end{equation}
In summary Eqs. (23), (37) and (38) provide the tools needed
to calculate the perpendicular field magnetoconductance. 

Let us now explicitly calculate the magnetoconductance. It is 
necessary to first evaluate $K_{B_{\perp}}^{(n)}({\mathbf r},{\mathbf r})$.
The integrals are performed in Appendix B and will be used in later
sections. For the present, more insight is gained by noticing
that $K_{B_{\perp}}^{(n)}({\mathbf r},{\mathbf r})$ has a natural interpretation
in terms of random walks. 

Imagine that $n$-sided polygons are drawn on a plane starting from
a fixed point ${\mathbf r}$. Let the probability density of drawing 
a polygon with vertices at ${\mathbf r}, \,{\mathbf r}_1, \,{\mathbf r}_2, 
\ldots,
\,{\mathbf r}_{n}, \,{\mathbf r}$ be given by the Gaussian expression
$ [ 2 \pi l_{e}^2 (n+1) ] K_{B=0}^{(0)} ({\mathbf r},{\mathbf r}_1)
K_{B=0}^{(0)}({\mathbf r}_1,{\mathbf r}_2) \ldots 
K_{B=0}^{(0)}({\mathbf r}_{n},{\mathbf r})$. The factor
$ 2 \pi l_{e}^2 (n+1)$ is included to normalize the distribution.
Let $a$ denote the directed area of the polygon (defined in
Appendix C). Then Eq. (37) may be rewritten as
\begin{equation}
K_{B_{\perp}}^{n} ({\mathbf r},{\mathbf r} )
= \frac{1}{2 \pi l_{e}^2 (n+1)} \int_{-\infty}^{+\infty} \!\!\!\!\!\! d a
\, P_{n+1}(a) \exp \left( 2 \pi i \frac{B_{\perp} a}{\phi_{{\rm sc}}}
\right).
\end{equation}
Here $P_{n}(a)$ is the probability distribution of the directed
area of an $n$-sided polygon drawn from the Gaussian polygon 
ensemble defined above. Eq. (39) shows that $K_{B_{\perp}}^{(n)}
({\mathbf r},{\mathbf r})$ is the Fourier transform of the distribution
of directed areas. The directed area distribution is discussed in 
Appendix C. For $n \gg 1$
\begin{equation}
P_{n}(a) = \frac{\pi}{2 n l_{e}^2} {\rm sech}^2 \left( 
\frac{\pi a}{n l_{e}^2} \right),
\end{equation}
and its Fourier transform is
\begin{equation}
K_{B_{\perp}}^{(n)}({\mathbf r},{\mathbf r}) = \frac{B_{\perp}}{2 \phi_{{\rm sc}}}
\left[ \sinh \left( \frac{n \pi B_{\perp} l_{e}^2 }{\phi_{{\rm sc}}}
\right) \right]^{-1}.
\end{equation}
Eq. (41) is derived in Appendix B and is shown to be sufficiently
accurate for our purposes for all $n$ provided $B_{\perp} \!\ll\! B_{e}$.
For further discussion of the directed area distribution see Appendix C.

Substitution of Eq. (41) in Eq. (38) leads to 
\begin{equation}
g_{{\rm WL}} = - \frac{e^2}{\hbar} \frac{W}{L} \sum_{n=1}^{\infty}
\frac{B_{\perp}}{2 B_e} \left[ \sinh \left( \frac{n B_{\perp}}{ 2
B_e} \right) \right]^{-1} \! \exp \left( - \frac{n \tau_{e}}{\tau_{\phi}}
\right).
\end{equation}
Since $B_{\perp} \! \ll\!  B_e$, the sum may be evaluated 
asymptotically.\footnote{The lower limit of the sum requires
delicate handling. Introduce a cutoff $N \gg 1$ but with
$N B_{\perp}/B_{e} \ll 1$ and $ N \tau_{e}/\tau_{\phi} \ll 1$.
The sum over $n$ from $1$ to $N$ is approximately $\ln N$ since the
summand is approximately $1/n$ over this range. The sum from $N$
to infinity may be turned into an integral in the limit $B_{\perp}/2
B_{e} \rightarrow 0$. The integral diverges logarithmically at its
lower limit. The divergence may be extracted by integration by
parts. This cancels the log dependence of the first sum on the
arbitrary cutoff $N$ and leaves a finite expression that can be 
expressed in terms of the digamma function using Eq. (43).}
Using an integral representation of the digamma function \cite{morse}
\begin{equation}
\psi (x) = \int_{0}^{\infty} d q \left( \frac{ e^{-q} }{q} -
\frac{ e^{-xq} }{1 - e^{-q}} \right) \;,
\end{equation}
we obtain
\begin{equation}
g_{{\rm WL}} = - \frac{1}{2 \pi^2} \frac{e^2}{\hbar} \frac{W}{L}
\left[ \ln \left( \frac{B_e}{B_{\perp}} \right) - \psi \left(
\frac{1}{2} + \frac{ B_{\phi} }{ B_{\perp} } \right) \right].
\end{equation}
Here $B_\phi$ is the field corresponding to a flux quantum through
the phase-breaking length, $B_\phi \equiv \phi_{\rm sc} / 4\pi l_\phi^2$.
This is the ideal weak-localization lineshape derived by
Hikami, Larkin and Nagaoka \cite{hikami} and shown as the dotted curves in 
Figs. 1 and 2. 

It is instructive to summarize the main points of the derivation.
First the conductance is expressed as a sum of closed random walks,
Eq. (38). Eq. (37) shows that the contribution of a fixed random
walk oscillates with the enclosed flux with a period $\phi_{{\rm sc}}$,
revealing the Aharonov-Bohm origin of the weak-localization 
magnetoresistance. The contribution of all $n$-bounce paths is 
the Fourier transform of their directed area distribution; it decays
exponentially for $B_{\perp} \gg \phi_{{\rm sc}}/n l_{e}^{2}$ [see 
Eq. (42)]. Summing over $n$ leads to the gentler decay described by 
Eq. (44). 

\section{Short Range Roughness}

In this section we assume that the distance over which the roughness
is correlated is short compared to the elastic mean free path
($L \ll l_e$). Under this circumstance the effect of turning on
an in-plane magnetic field is comparatively straightforward:
it enhances the dephasing rate and hence produces ``homogeneous
broadening'' of the weak-localization lineshape.

For simplicity we first assume that the applied field lies
in the plane of the interface. According to the analysis of
Section II, the interface electrons see a uniform in-plane 
field as a random perpendicular field. The typical strength
of the random field is $B_g \sim B_{\parallel} \Delta/L$ [see
Eqs. (5) and (12)]. We assume that the random magnetic field
is weak compared to impurity scattering ($B_g \ll B_e$), a condition
that is well satisfied in practice. As
in the previous section, the only effect of the magnetic field
in this limit is to multiply the Green function by a phase
factor, so the zeroth-order Cooperon becomes
\begin{equation}
C_{B_{\parallel}}^{(0)} ({\mathbf r},{\mathbf r}') = C_{B=0}^{(0)} 
({\mathbf r},{\mathbf r}') \exp \left( i \frac{2e}{\hbar} 
\int_{{\mathbf r}'}^{{\mathbf r}} d {\mathbf l}. {\mathbf A}_g \right).
\end{equation}
The Cooperon is determined in principle by Eqs. (17) and
(45); our objective is to average it over the interface
roughness.

To this end it is convenient to formally solve Eq. (17) by iteration:
\begin{eqnarray}
C_{B_{\parallel}}({\mathbf r},{\mathbf r}) & = & \sum_{n=0}^{\infty}
C_{B_{\parallel}}^{(n)} ({\mathbf r},{\mathbf r}) ; \nonumber \\
C_{B_{\parallel}}^{(n)} ({\mathbf r},{\mathbf r}) & = &
\int d {\mathbf r}_1 \ldots d {\mathbf r}_{n} C_{B=0}^{(0)} ({\mathbf r},
{\mathbf r}_1) \ldots \nonumber \\
& & \ldots C_{B=0}^{(0)} ({\mathbf r}_{n},{\mathbf r}) 
\exp \left( i 2 \pi \frac{\phi_g}{\phi_{{\rm sc}}} \right);
\nonumber \\
\phi_g & = & \int_{{\mathbf r}}^{{\mathbf r}} d 
{\mathbf l} \cdot {\mathbf A}_g .
\end{eqnarray}
The integral in Berry's phase factor, $\phi_g$, 
is to be evaluated around the perimeter of the polygon 
obtained by joining
${\mathbf r},{\mathbf r}_1, \ldots, {\mathbf r}_{n},{\mathbf r}$ in 
that order. We average the Berry phase factor using the
gauge given by Eq. (8) and assuming that ${\cal Z}_g (x,y)$
is a Gaussian random surface with statistics given by Eq. (12).
To avoid digression we shall return to the evaluation of this
average below. For the moment we give the result for $L \!\ll\! l_e$:
\begin{eqnarray}
 & &
\left\langle \exp \left( i 2 \pi \frac{\phi_g}{\phi_{{\rm sc}}} \right) 
\right\rangle_{{\rm rough}} \approx
\nonumber \\
 & & 
\exp \left( - 2 \sqrt{\pi} \frac{e^2}{\hbar^2} B_{\parallel}^2 
\Delta^2 L \left[ \frac{(y - y_1)^2}{|{\mathbf r} - {\mathbf r}_1|} +
\ldots + \frac{ (y_{n} - y)^2 }{ |{\mathbf r}_{n} - {\mathbf r}| } 
\right] \right). \nonumber \\
 & & 
\end{eqnarray}
$ $ From Eqs. (46) and (47) we obtain the average Cooperon,
\begin{eqnarray}
{\cal C}_{B_{\parallel}} ({\mathbf r},{\mathbf r}) & = & \sum_{n=0}^{\infty}
{\cal C}_{B_{\parallel}}^{(n)} ({\mathbf r},{\mathbf r}) ; \\
{\cal C}_{B_{\parallel}}^{(n)} ({\mathbf r},{\mathbf r}) & = & \int
d {\mathbf r}_1 \ldots d {\mathbf r}_{n} 
{\cal C}_{B_{\parallel}}^{(0)} ({\mathbf r},{\mathbf r}_1) \ldots
{\cal C}_{B_{\parallel}}^{(0)} ({\mathbf r}_{n},{\mathbf r}) ; \nonumber \\
{\cal C}_{B_{\parallel}}^{(0)} ({\mathbf r},{\mathbf r}') & = & C_{B=0}^{(0)}
({\mathbf r},{\mathbf r}') \exp \left( \!- 2 \sqrt{\pi} \frac{e^2}{\hbar^2} 
B_{\parallel}^2 \Delta^2 L \frac{(y - y')^2}{| {\mathbf r} - {\mathbf r}' |}
\right). \nonumber 
\end{eqnarray}
For brevity $ \langle C_{B_{\parallel}} ({\mathbf r},{\mathbf r}) 
\rangle_{{\rm rough}} $ is written as ${\cal C}_{B_{\parallel}} 
({\mathbf r},{\mathbf r})$. 

Evidently Eq. (48) is the formal iterative solution to the integral
equation
\begin{equation}
{\cal C}_{B_{\parallel}} ({\mathbf r},{\mathbf r}') = 
{\cal C}_{B_{\parallel}}^{(0)} ({\mathbf r},{\mathbf r}') +
\int d {\mathbf r}'' {\cal C}_{B_{\parallel}}^{(0)} ({\mathbf r},{\mathbf r}'')
{\cal C}_{B_{\parallel}}({\mathbf r}'',{\mathbf r}').
\end{equation}
In principle the roughness averaged Cooperon is determined
by Eq. (49) with ${\cal C}_{B_{\parallel}}^{(0)} ({\mathbf r},{\mathbf r}')$
given by Eqs. (21) and (48). The expression for 
${\cal C}^{(0)}_{B_{\parallel}} ({\mathbf r}, {\mathbf r}')$
shows that after averaging
over roughness, the effect of the in-plane field is to multiply
the zeroth-order Cooperon by an exponential factor.

To solve Eq. (49) it is most convenient to look for eigenfunctions
of $C_{B_{\parallel}}^{(0)}$ which obey
\begin{eqnarray}
& & 
\int \! d {\mathbf r}' C_{B=0}^{(0)} ({\mathbf r},{\mathbf r}') 
\exp \left( \!- 2 \sqrt{\pi} \frac{e^2}{\hbar^2} B_{\parallel}^2
\Delta^2 L \frac{ (y - y')^2 }{|{\mathbf r} - {\mathbf r}'|} \right)
Q_{\lambda}({\mathbf r}') \nonumber \\
& & 
= \lambda Q_{\lambda}({\mathbf r}).
\end{eqnarray}
${\cal C}_{B_{\parallel}}$ may be expanded 
in terms of these eigenfunctions as in Eq. (20). 
We are interested in eigenfunctions that vary
slowly compared to $l_e$. For these eigenfunctions Eq. (50) may be
transformed into a diffusion equation by expanding $Q_{\lambda}$
and the exponential factor in powers of ${\mathbf r}-{\mathbf r}'$. 
Keeping the leading terms (second order in $Q_{\lambda}$ and
first-order in the exponential factor) we obtain
\begin{equation}
\left( 1 + D \tau_e \nabla^2 \right) Q_{\lambda}({\mathbf r})
= \left( \lambda + \frac{\tau_e}{\tau_{\phi}}
+ \frac{\tau_e}{\tau_{\parallel}} \right) Q_{\lambda}
({\mathbf r}).
\end{equation}
In Eq. (51) we have put a dephasing term $\tau_e/\tau_{\phi}$
by hand. The in-plane field is responsible for the $\tau_e/
\tau_{\parallel}$ term. Here $\tau_{\parallel}$ is defined by
\begin{eqnarray}
\frac{\tau_e}{\tau_{\parallel}} & = & 2 \sqrt{\pi} 
\frac{e^2}{\hbar^2} B_{\parallel}^2 \Delta^2 L 
\int d {\mathbf r}' C_{B=0}^{(0)} ({\mathbf r},{\mathbf r}') 
\frac{(y-y')^2}{|{\mathbf r}-{\mathbf r}'|} \nonumber \\
 & \approx & \sqrt{\pi} \frac{e^2}{\hbar^2} B_{\parallel}^2 
\Delta^2 L l_e.
\end{eqnarray}
Eq. (51) shows that, upon averaging, the only effect of an
in-plane magnetic field on the long-distance behavior of the
Cooperon, and hence on the conductance, is to enhance the
dephasing rate by 
\begin{equation}
\frac{1}{\tau_{\phi}} \rightarrow \frac{1}{\tau_{\phi}} +
\frac{1}{\tau_{\parallel}}
\end{equation}
where $\tau_{\parallel}$ is given by Eq. (52).

The form of $\tau_{\parallel}^{-1}$ in Eq. (52) can be obtained
from the following simple argument. We wish to find the typical phase 
associated with an n-sided random polygon due to the random vector
potential. Consider a single side of the polygon: it has typical
length $l_e$, and the vector potential is approximately constant
over segments of length $L$ but of either 
sign---$A \sim \pm B_{\parallel}\Delta$. 
Then the typical phase accumulated on
this side is $\text{rms} [ (2e/\hbar) \int {\bf A} \cdot d{\bf l} ] \sim$
$ (2e/\hbar) \sqrt{l_e/L}) B_{\parallel} \Delta L$. Since the phase
along each side is uncorrelated, the typical total phase is
$\text{rms} [ \phi_{\rm n-gon} ] \sim$ $\sqrt{n} (2e/\hbar) \sqrt{l_e/L}) 
B_{\parallel} \Delta L $. The contribution of n-gons will be cutoff
when this phase is of order 1, giving the condition for 
$\tau_{\parallel}^{-1}$. This gives the condition for $\tau_{\parallel}^{-1}$
if we set $ n \rightarrow \tau_{\parallel}/\tau_e$. We find 
$\tau_e/\tau_{\parallel} \sim$ $4(e/\hbar)^2 B_{\parallel}^2 \Delta^2 L l_e$,
the same form as Eq. (52).

Now we indicate how to generalize this analysis when the applied
magnetic field has a non-zero perpendicular component. Assuming
this component is weak compared to impurity scattering $(B_{\perp}
\ll B_e$), an Aharonov-Bohm phase factor must be included in
Eqs. (45) and (46). After averaging over interface roughness
${\cal C}_{B_{\parallel},B_{\perp}}$ obeys Eq. (49) with
\begin{eqnarray}
{\cal C}_{B_{\parallel},B_{\perp}}^{(0)} ({\mathbf r},{\mathbf r}') & = & 
C_{B=0}^{(0)} ({\mathbf r},{\mathbf r}') \exp \left( i \frac{2e}{\hbar}
\int_{{\mathbf r}'}^{{\mathbf r}} d {\mathbf l}.{\mathbf A} \right)
\\
& &
\times \exp \left( - 2 \sqrt{\pi} \frac{e^2}{\hbar^2} B_{\parallel}^2
\Delta^2 L \frac{ (y-y')^2 }{ |{\mathbf r} - {\mathbf r}'| } \right).
\nonumber
\end{eqnarray}
Keeping track of the phase factor, we find that the slowly 
varying eigenfunctions of ${\cal C}_{B_{\parallel},B_{\perp}}^{(0)}$
obey Eq. (33) with the substitution given in Eq. (53).

Hence we arrive at the central result of this section. 
{\it For short-range correlated roughness, application of an in-plane
field increases the dephasing rate according to Eq. (53) but does
not otherwise alter the weak-localization lineshape.} The dependence
of the conductance on the perpendicular field is still given by
Eq. (44) but with 
\begin{equation}
B_{\phi} \rightarrow B_{\phi} + \pi^{3/2} 
\frac{e B_{\parallel}^2}{h} \frac{\Delta^2 L}{l_e}.
\end{equation}
Since the form of the lineshape is not changed, this effect
is analogous to {\it homogeneous} broadening in magnetic resonance
and atomic physics. 

The idea that an in-plane field produces homogeneous broadening of the
line was anticipated by the authors of Refs.~\onlinecite{bob1,bob2}.  Our 
result differs in two important respects: First, we find that homogeneous
broadening occurs only if the roughness is correlated over short distances
($L \ll l_e$); and second, the dependence of $\tau_{\parallel}$ on the
interface roughness parameters that we derive [Eq. (52)] is quite
different from that conjectured in Ref.~\onlinecite{bob2}.

Fig. 6 shows that Eqs. (44) and (55) provide a good fit to data with
essentially no adjustable parameters. More details are given in Section
VI.

To complete the analysis we must now average Berry's phase factor
$\exp ( i 2 \pi \phi_g /\phi_{{\rm sc}} ) $ over interface roughness
to obtain Eq. (47). Recall that the distribution of a linear 
combination of correlated Gaussian random variables is also
Gaussian. The distribution of $\phi_g$ is therefore Gaussian.
Evidently it has zero mean since we have assumed $\langle
{\cal Z}_g(x,y) \rangle_{{\rm rough}} = 0$. For a Gaussian
random variable, $s$, with zero mean it is easy to show
\begin{equation}
\langle \exp ( i \xi s ) \rangle =
\exp \left( - \case{1}{2} \xi^2 \langle s^2 \rangle \right).
\end{equation}
Hence averaging Berry's phase factor reduces to the calculation
of the variance of the phase. Explicitly we must calculate
\begin{equation}
\langle \phi_g^2 \rangle_{{\rm rough}} =
\sum_{i=1}^{n+1} \sum_{j=1}^{n+1} \Bigl\langle 
\int \! d {\mathbf l}_i \cdot {\mathbf A}_g  
\int \! d {\mathbf l}_j \cdot {\mathbf A}_g \Bigr\rangle_{{\rm rough}}.
\end{equation}
Here $ \int d {\mathbf l}_i \cdot {\mathbf A}_g $ denotes the line integral
along the edge joining $ {\mathbf r}_{i-1}$ to ${\mathbf r}_i$ (with
the understanding that ${\mathbf r}_{0} = {\mathbf r}_{n+1} = {\mathbf r}$).

Let us first evaluate the diagonal terms in Eq. (57). Let the
$i^{{\rm th}}$ edge make an angle $\theta$ with the $x$-axis
and let $s$ be the length parameter along that edge. Using
Eq. (8) for ${\mathbf A}_g$ and Eq. (12) for the correlation of 
${\cal Z}_g (x,y)$ we obtain
\begin{eqnarray}
 & & \Bigl\langle ( \int d {\mathbf l}_i \cdot {\mathbf A}_g )^2 
     \Bigr\rangle_{{\rm rough}}
= \nonumber \\
 & & B_{\parallel}^2 \Delta^2 \sin^2 \theta 
\int_{0}^{| {\mathbf r}_i - {\mathbf r}_{i-1} |} \!\!\!\!\!\! d s
\int_{0}^{| {\mathbf r}_i - {\mathbf r}_{i-1} |} \!\!\!\!\!\! d s'
\exp \biggl[ - \left( \frac{s-s'}{L} \right)^2 \biggr]
\nonumber \\
 & & \approx \sqrt{\pi} B_{\parallel}^2 \Delta^2 L 
\frac{ (y_{i} - y_{i-1})^2 }{ | {\mathbf r}_i - {\mathbf r}_{i-1} | }.
\end{eqnarray}
We have assumed that the edge is much longer than the correlation
length $L$. Since the length of a typical edge is $l_e$, this
is justified in the short range limit $L \ll l_e$. We have also
used $|{\mathbf r}_i - {\mathbf r}_{i-1} | \sin \theta = y_i - y_{i-1}$.

Evidently the only important off diagonal terms in Eq. (57) are those for
which the two edges pass within a correlation length of each other. The
contribution of such edge pairs to the sum will be smaller than the
diagonal terms, typically by a factor of $L/l_e$ (except in the rare
circumstance that the two edges meet at a glancing angle and overlap for a
considerable part of their length). Furthermore, the number of
contributing off-diagonal terms is comparable to $n$, the number of
diagonal terms; hence the contribution of the off-diagonal terms may be
safely neglected in the short range limit $L \ll l_e$.

That the number of important off-diagonal terms is comparable to $n$ can
be established by an argument commonly used in polymer physics
\cite{edwards}. Suppose that line segments of typical length $l_e$ are
drawn at random on a plane with a density $ \rho $ segments per unit area.
If a fresh segment of length $l_e$ is now drawn on  this plane, typically
it will suffer $\sim \rho l_e^2$ intersections.  Working in units where
$l_e = 1$, the area of a typical polygon in Eq. (46) is $n$ and hence the
density of edges $\rho \sim n/n = 1$.  Each edge therefore typically
encounters one intersection and the total number of intersections is
comparable to $n$.

Adding the diagonal contributions [Eq. (58)] and neglecting
the off-diagonal ones we obtain
\begin{equation}
\langle \phi_g^2 \rangle_{{\rm rough}} \approx \sqrt{\pi} 
B_{\parallel}^2 \Delta^2 L \left[ \frac{ (y_1 - y )^2 }{ |
{\mathbf r}_1 - {\mathbf r} | } + \ldots + \frac{ (y - y_{n})^2 }{ |
{\mathbf r} - {\mathbf r}_{n} | } \right].
\end{equation}
Use of Eqs. (56) and (59) finally leads to Eq. (47).

\section{Long Range Roughness}

We now consider the roughness correlation length to be long compared to
the mean free path ($L \gg l_e$). This circumstance is more difficult to
analyze; hence, the discussion is limited to the circumstance that the
in-plane magnetic field is weak in a sense made precise below. Roughly it
is assumed that the random Berry phase is small for all contributing
paths.\footnote{This is more stringent than the condition $B_{g} \sim
B_{\parallel} \Delta/L \ll B_e$ imposed in the previous section [see
discussion following Eq. (45)].} The experiments of
Refs.~\onlinecite{bob1,bob2,albert} meet this condition for most part. For
simplicity, we first assume the applied field lies in the plane of the
interface, $B_{\perp} = 0$ (Section Va). In the next subsection no
restriction is placed on $B_{\perp}$ (other than $B_{\perp} \ll B_e$).

\subsection{Zero Perpendicular Field Applied}

Since $B_{g} \ll B_e$ is automatically ensured
by the weak field condition on $B_{\parallel}$ that
we will impose below, the Cooperon is given by
Eq. (45). Since the Berry phase factor varies
on the scale of $L \gg l_e$, we may work with
$K$ instead of $C$ and write
\begin{equation}
K^{(0)}_{B_{\parallel}} ({\mathbf r},{\mathbf r}') =
K^{(0)}_{B=0} ({\mathbf r},{\mathbf r}') 
\exp \left( i \frac{2 e}{\hbar} \int_{{\mathbf r}'}^{{\mathbf r}} d 
{\mathbf l}.{\mathbf A}_{g} \right)
\end{equation}
in place of Eq. (45) and 
\begin{eqnarray}
K_{B_{\parallel}} ({\mathbf r},{\mathbf r}) & = &
\sum_{n=0}^{\infty} K^{(n)}_{B_{\parallel}} ({\mathbf r},
{\mathbf r}); \nonumber \\
K_{B_{\parallel}}^{(n)} ({\mathbf r},{\mathbf r}) & = & 
\int d {\mathbf r}_1 \ldots d {\mathbf r}_{n} K^{(0)}_{B=0} ({\mathbf r},
{\mathbf r}_1) \ldots \nonumber \\
 & & \ldots K^{(0)}_{B=0} ({\mathbf r}_{n}, {\mathbf r}) 
\exp \left( i 2 \pi \frac{ \phi_{g} }{ \phi_{{\rm sc}} } \right);
\nonumber \\
\phi_{g} & = & \int_{{\mathbf r}}^{{\mathbf r}} d {\mathbf l} \cdot 
 {\mathbf A}_{g}
\end{eqnarray}
in place of Eq. (46). Note that we could not make this replacement
in the previous section since the Berry phase fluctuated rapidly 
on the scale of $l_e$. Our objective now is to average 
$K_{B_{\parallel}}$ over interface roughness and then use 
Eq. (27) to evaluate the conductance.

The Berry phase factor is a Gaussian random variable so
it is easy to perform the average. Using Eq. (8) for the 
vector potential, Eq. (12) for its statistics and Eq. (56),
we obtain
\begin{eqnarray}
{\cal K}^{(n)}_{B_{\parallel}} ({\mathbf r},{\mathbf r}) 
& = & \int d {\mathbf r}_{1} \ldots d {\mathbf r}_{n} 
K^{(0)}_{B=0} ({\mathbf r},{\mathbf r}_1) \ldots 
\\
& & \ldots 
K^{(0)}_{B=0} ({\mathbf r}_{n},{\mathbf r})  
\exp \left[ - V( {\mathbf r}, {\mathbf r}_1, \ldots, {\mathbf r}_{n} )
\right] \nonumber 
\end{eqnarray}
with 
\begin{eqnarray}
\lefteqn{ V({\mathbf r},{\mathbf r}_1, \ldots, {\mathbf r}_{n}) = } \\
& & \quad \frac{ 4 e^2 B_{\parallel}^2 \Delta^2 }{ \hbar^2 }
\sum_{j=1}^{n} \sum_{k=1}^{n}
(y_{j} - y_{j-1})(y_{k} - y_{k-1})
\nonumber \\
 & & \quad \times
\exp \left[ - \frac{1}{4 L^2} ( {\mathbf r}_j + {\mathbf r}_{j-1}
- {\mathbf r}_{k} - {\mathbf r}_{k-1} )^2 \right]. \nonumber
\end{eqnarray}
Here ${\cal K}^{(n)} = \langle K^{(n)} \rangle_{{\rm rough}}$
and $ {\mathbf r}_{0} = {\mathbf r}_{n+1} = {\mathbf r}$.
${\cal K}^{(n)}$ is thus an integral over $(n+1)$-bounce closed
random walks just as it is at zero field but the weight 
of each polygon is no longer Gaussian. The links of the
polygon now ``interact'', and the interaction, given by
Eq. (63), is anisotropic and long-ranged. 

${\cal K}^{(n)}$ is difficult to calculate.  For weak in-plane magnetic fields
it is sufficient to analyze the interaction perturbatively.  To first order
\begin{eqnarray}
\delta {\cal K}^{(n)}_{B_{\parallel}} ({\mathbf r}, {\mathbf r})
& \equiv & 
{\cal K}^{(n)}_{B_{\parallel}} ({\mathbf r}, {\mathbf r})
- K^{(n)}_{B=0} ({\mathbf r},{\mathbf r}) 
\\
 & = & 
- \int d {\mathbf r}_{1} \ldots d {\mathbf r}_{n} 
K^{(0)}_{B=0} ({\mathbf r}, {\mathbf r}_{1}) \ldots
\nonumber \\
 & & 
\ldots K^{(0)}_{B=0} ({\mathbf r}_{n}, {\mathbf r}) 
V({\mathbf r},{\mathbf r}_{1}, \ldots, {\mathbf r}_{n}).
\nonumber
\end{eqnarray}
$V({\mathbf r}, {\mathbf r}_1, \ldots, {\mathbf r}_{n})$
is a sum over pairs of links according to Eq. (63).
To illustrate the evaluation of Eq. (64) focus on
the particular term for which $j=1$ and $ k = 3, 4,
\ldots, n$. It is convenient to perform the 
integrals over the end points of the $j^{{\rm th}}$
and $k^{{\rm th}}$ links last. Integrating over
the remaining intermediate points 
yields
\begin{eqnarray}
\delta {\cal K}^{(n)}_{B_{\parallel}} ({\mathbf r},{\mathbf r})
& = &
\frac{1}{\pi^2} \int d {\mathbf r} d {\mathbf r}_{1} d {\mathbf r}_{k-1}
d {\mathbf r}_{k} 
\\
& & \times
\frac{1}{(n+1-k)(k-2)} (y_{1}-y)(y_{k} - y_{k-1})  
\nonumber \\
& & \times
\exp \left[ - \frac{ ({\mathbf r} - {\mathbf r}_{k})^2 }{ 2 l^{2}_{e} (n+1 - k) }
- \frac{ ( {\mathbf r}_{k-1} - {\mathbf r}_{1} )^2 }{ 2 l^{2}_{e} (k - 2) } \right]
\nonumber \\
& & \times
\exp \left[ - \frac{ ( {\mathbf r}_{k} - {\mathbf r}_{k-1} )^2 }{ 2 l^{2}_{e} }
- \frac{ ( {\mathbf r}_{1} - {\mathbf r} )^2 }{ 2 l^{2}_{e} } \right]
\nonumber \\
& & \times
\exp \left[ \frac{1}{4 L^2} ( {\mathbf r} + {\mathbf r}_1 - {\mathbf r}_{k-1}
- {\mathbf r}_{k} )^2 \right]
\nonumber
\end{eqnarray}
This is a low-dimensional Gaussian integral and can be
explicitly evaluated. 

Proceeding in this manner we obtain
\begin{eqnarray}
\delta {\cal K}^{(n)}_{B_{\parallel}} ({\mathbf r},{\mathbf r}) & = &
\Bigl( \frac{e B_{\parallel} \Delta}{ \hbar } \Bigr)^2
\! [ f_{1}(n, d) + f_{2}(n, d) + f_{3}(n, d) ];
\nonumber \\
f_{1}(n, d) & \approx & \frac{ n(1 + n d) }{16 d^2} \sum_{k=3}^{n}
\frac{1}{ ( k^2 - n k - [n/4d] )^2 } ;
\nonumber \\
f_{2}(n, d) & \approx & \frac{1 + n d}{ n } ;
\nonumber \\
f_{3}(n, d) & \approx & 1
\end{eqnarray}
with $ d = l_{e}^{2}/(2 L^2)$. The three contributions to
${\cal K}^{(n)}$ arise from the interaction of disconnected
links, adjacent links, and from the self-interaction of links,
respectively. We have used $n \gg 1$ and $d \ll 1$ to simplify
Eq. (66). This is justified because weak-localization is 
dominated by long paths and we are concerned with long-range
correlated roughness in this section.

Using Eq. (27) for the conductance and Eq. (66) for 
$\delta {\cal K}^{(n)}$, after considerable simplification
we obtain the parallel field magnetoconductance
\begin{equation}
\delta g = - \frac{e^2}{\hbar} \frac{1}{\phi_{{\rm sc}}^2} 
( B_{\parallel} \Delta L )^2 f( \alpha )
\end{equation}
with $ \alpha = l_{\phi}^2/L^2 $ and
\begin{eqnarray}
f( \alpha ) & = & \alpha \int_{0}^{\infty} d x e^{-x}
\\
& & \times
\left( 1 + \frac{1}{2 [ x \alpha (1 + x \alpha) ]^{1/2}}
\ln \frac{ \sqrt{1 + x \alpha} - \sqrt{ x \alpha } }{
\sqrt{ 1 + x \alpha } + \sqrt{ x \alpha } } \right).
\nonumber
\end{eqnarray}
{\it It is striking that $ \delta g$ is independent of the mean
free path $ l_{e} $.} In appropriate units it is a product
of two factors: the flux through an area $ \Delta L $, 
determined entirely by the geometry of the rough interface,
and $ f(\alpha)$. $f(\alpha)$ describes the crossover as
$ l_{\phi}$ is varied relative to $L$. It is plotted in 
Fig. 4 and has the asymptotic behavior
\begin{eqnarray}
f( \alpha ) & \approx & \alpha \hspace{3mm} {\rm for}
\hspace{3mm} \alpha \gg 1 ;
\nonumber \\
 & \approx & \frac{2}{3} \alpha^2 \hspace{3mm} {\rm for}
\hspace{3mm} \alpha \ll 1.
\nonumber \\
\end{eqnarray}  

\begin{figure}[t]
\begin{center}
\leavevmode
\epsfxsize=3.0in \epsfbox{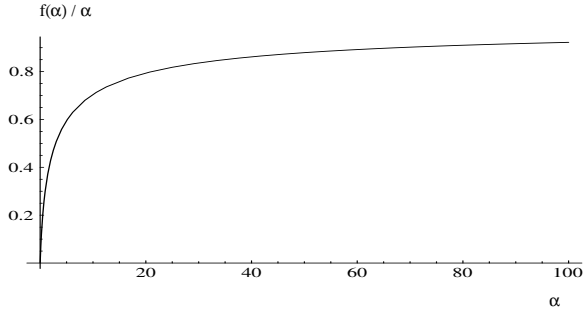}
\end{center}
\caption{Plot of $f(\alpha)/\alpha$ as a function of
$\alpha = l_{\phi}^{2}/L^2$. $f(\alpha)$ describes the dependence
of the in-plane magnetoresistance, at zero perpendicular field, on
the roughness correlation length, $L$ (see Eq. 67). }
\end{figure}

Substituting Eq. (69) in Eq. (67) gives the asymptotic behavior
of the parallel field magnetoconductance
\begin{eqnarray}
\delta g & \approx & - \frac{e^2}{\hbar} \frac{ B_{\parallel}^2
\Delta^2 l_{\phi}^{2} }{ \phi_{{\rm sc}}^2 } \hspace{3mm} 
{\rm for} \hspace{3mm} l_{\phi} \gg L;
\nonumber \\
 & \approx & - \frac{2}{3} \frac{e^2}{\hbar} \frac{ B_{\parallel}^2
\Delta^2 l_{\phi}^{4} }{ \phi_{{\rm sc}}^{2} L^{2} } \hspace{3mm}
{\rm for} \hspace{3mm} L \gg l_{\phi}.
\end{eqnarray}
These formulae have a simple interpretation. The random magnetic
field has a typical magnitude $B_{\parallel} \Delta / L$ and is
correlated over a distance $L$. For $L \gg l_{\phi}$, the typical
flux through an area $l_{\phi}^2$ is therefore $ (B_{\parallel} 
\Delta / L) l_{\phi}^2 $; for $ L \ll l_{\phi}$, it is $B_{\parallel}
\Delta l_{\phi}$ (see 
footnote\footnote{The area $l_{\phi}^2$ can be broken up
into $ l_{\phi}^2/L^2 $ correlated squares of area $L^2$. The
typical flux through each square is $(B_{\parallel} \Delta / L) L^2$.
Since it can be of either sign, the flux through the total area 
$l_{\phi}^2$ grows as the square root of the number of squares.}).
Eq. (70) then shows that $ \delta g$ (in units of $ e^2/h $) is the
square of the flux through a phase-coherent region in units of the
flux quantum.

\subsection{Perpendicular Field Applied}

The calculation with a perpendicular field applied closely
parallels the calculation at $B_{\perp} = 0$. We write
\begin{equation}
K_{B_{\perp}, B_{\parallel}}^{(0)} ({\mathbf r},{\mathbf r}') = 
K_{B=0}^{(0)} ({\mathbf r}, {\mathbf r}') \exp \left[ i \frac{ 2 e }{\hbar}
\int_{{\mathbf r}'}^{{\mathbf r}} d {\mathbf l} \cdot \left( {\mathbf A}_{g}
+ {\mathbf A} \right) \right]
\end{equation}
in place of Eq. (60) and 
\begin{eqnarray}
K_{B_{\perp},B_{\parallel}} ({\mathbf r},{\mathbf r}) & = &
\sum_{n=0}^{\infty} K_{B_{\perp}, B_{\parallel}}^{(n)}
({\mathbf r}, {\mathbf r}) ; \nonumber \\
K_{B_{\perp},B_{\parallel}}^{(n)} ({\mathbf r}, {\mathbf r}) & = & 
\int d {\mathbf r}_{1} \ldots d {\mathbf r}_{n} K_{B=0}^{(0)}
({\mathbf r}, {\mathbf r}_{1}) \nonumber \\
 & & \ldots K_{B=0}^{(0)}({\mathbf r}_{n}, {\mathbf r}) \ldots
\nonumber \\
& & \ldots \exp \left( i 2 \pi \frac{ \phi_{g} }{ \phi_{{\rm sc}} }
\right) \exp \left( i 2 \pi \frac{ \phi_{\perp} }{ \phi_{{\rm sc}} }
\right); \nonumber \\
\phi_{g} & = & \int_{{\mathbf r}}^{{\mathbf r}} d {\mathbf l} \cdot {\mathbf A}_{g} ;
\nonumber \\
\phi_{\perp} & = & \int_{{\mathbf r}}^{{\mathbf r}} d {\mathbf l} \cdot {\mathbf A}
\end{eqnarray}
in place of Eq. (61). 
Here $\phi_{\perp}$ is the Aharonov-Bohm 
flux through the polygon that goes from
${\mathbf r}$ to ${\mathbf r}$ via ${\mathbf r}_{1} \ldots {\mathbf r}_{n}$
due to the applied perpendicular field; $ \phi_{g}$ is the
Berry phase around the same polygon.

The Berry phase factor is a Gaussian random variable. Performing
the average as in the previous subsection yields
\begin{eqnarray}
{\cal K}_{B_{\perp},B_{\parallel}}^{(n)} ({\mathbf r}, {\mathbf r})
& = & \int d {\mathbf r}_{1} \ldots d {\mathbf r}_{n} K_{B=0}^{(0)} 
( {\mathbf r}, {\mathbf r}_{1} ) \ldots \nonumber \\
 & & \ldots K_{B=0}^{(0)} ({\mathbf r}_{n}, {\mathbf r}) 
\exp [ - V({\mathbf r}, {\mathbf r}_{1}, \ldots, {\mathbf r}_{n}) ]
\nonumber \\
& &
\times \exp \left( i 2 \pi \frac{ \phi_{\perp} }{ \phi_{{\rm sc}} }
\right).
\end{eqnarray}
Here ${\cal K}^{(n)} = \langle K^{(n)} \rangle_{{\rm rough}}$
and $V$ is given by Eq. (63). As in Section IIIB, ${\cal K}^{(n)}$
is the Fourier transform of the directed area distribution of 
$n$-sided polygons; but the weight of the polygons is no longer
Gaussian. The links of the polygons now ``interact'' and the
interaction given by Eq. (63) is anisotropic and long-ranged.

For weak in-plane fields a full analysis of ${\cal K}^{(n)}$
is not needed. It is sufficient to analyze the interaction
perturbatively. To first order
\begin{eqnarray}
\delta {\cal K}^{(n)}_{B_{\perp},B_{\parallel}} ({\mathbf r}, {\mathbf r}) 
& \equiv &
{\cal K}^{(n)}_{B_{\perp},B_{\parallel}} ({\mathbf r}, {\mathbf r})
- K^{(n)}_{B_{\perp}} ({\mathbf r}, {\mathbf r}) 
\nonumber \\
 & = & - \int d {\mathbf r}_{1} \ldots d {\mathbf r}_{n} K^{(0)}_{B=0} 
({\mathbf r},{\mathbf r}_{1}) \nonumber \\
 & & \ldots K^{(0)}_{B=0} ({\mathbf r}_{n},{\mathbf r})
 V({\mathbf r}, \ldots, {\mathbf r}_{n})
\nonumber \\
 & & \times  
\exp \left( i \frac{2 \pi \phi_{\perp} }{ \phi_{{\rm sc}}}
\right). 
\end{eqnarray}
$V({\mathbf r}, {\mathbf r}_{1}, \dots, {\mathbf r}_{n})$ is a sum over
pairs of links according to Eq. (63). To evaluate Eq. (74) it is
convenient to focus on the particular term corresponding to the
$j^{{\rm th}}$ and $k^{{\rm th}}$ links. As in the corresponding
evaluation of Eq. (64), it is convenient to perform the integrals
over the end points of the $j^{{\rm th}}$ and $k^{{\rm th}}$ links
last. The integral over the other intermediate points is facilitated
by Eqs. (B10, B15) of Appendix B. The expression that results is a
low-dimensional Gaussian integral over the end points of links $j$
and $k$ that can be explicitly computed.

Proceeding in this manner we obtain
\begin{eqnarray}
\delta {\cal K}^{(n)}_{B_{\perp},B_{\parallel}} ({\mathbf r}, {\mathbf r} )
& = & \left( \frac{e B_{\parallel} \Delta}{ \hbar } \right)^2
[ f_{1} + f_{2} - f_{3}];
\nonumber \\
f_{1} (n, \gamma, d) & \approx & \frac{n \gamma}{\pi} \sum_{k=3}^{n}
\bigl\{ \coth[ k \sqrt{\gamma} ] \coth[ (n - k) \sqrt{\gamma} ] - 1
\nonumber \\
 & + &
\frac{d}{\sqrt{\gamma}} \coth [k \sqrt{\gamma}] + \frac{d}{\sqrt{\gamma}} 
\coth[ (n - k) \sqrt{\gamma} ] \bigr\}
\nonumber \\
 & \times & \bigl\{ \sinh [ k \sqrt{\gamma} ] \sinh [ (n - k) \sqrt{\gamma} ]
\bigr\}^{-1}
\nonumber \\
 & \times & \bigl( \coth [ k \sqrt{ \gamma } ] + \coth [ (n - k)
\sqrt{ \gamma} ] + 4 \frac{ d }{ \sqrt{ \gamma } } \bigr)^{-2}
\nonumber \\
f_{2} (n, \gamma, d) & \approx & \frac{2 n}{ \pi} \frac{ \sqrt{ \gamma } 
}{ \sinh [n \sqrt{ \gamma }] } \{ \sqrt{ \gamma } \coth 
[ n \sqrt{ \gamma } ] + d \}
\nonumber \\
f_{3} (n,\gamma, d) & \approx & \frac{n}{\pi} \sqrt{\gamma} 
\frac{1}{\sinh [ n \sqrt{ \gamma } ]}
\end{eqnarray}
where $\gamma \equiv (B_{\perp}/2 B_e)^2 $ is a measure of 
the dynamical strength of the applied perpendicular field
and we have taken $ n \gg 1$, $\gamma \ll 1 $ and $ d \ll 1$.
This is justified because weak-localization is dominated by
long paths, the applied field is dynamically weak, and we are
concerned with long-ranged roughness in this section. We have
also taken $ n \gamma \ll 1$, justifiable {\em a posteriori}, 
because we find $\delta {\cal K}^{(n)}$ decays exponentially
for $ n \gg 1/\sqrt{\gamma} $. 

Using Eq. (27) for the conductance and Eq. (75) for $\delta
{\cal K}^{(n)}_{B_{\perp}, B_{\parallel}}$, after considerable
simplification, we obtain a lengthy expression for the
parallel field magnetoconductance,
\begin{equation}
\delta g = -2 \frac{e^2}{\hbar} \left( \frac{ B_{\parallel} \Delta
L}{ \phi_{{\rm sc}} } \right)^2 u ( \eta, B_{\phi}/B_{\perp} ).
\end{equation}
Here $ \eta \equiv \phi_{{\rm sc}}/ 2 \pi B_{\perp} L^2 $ and
\begin{equation}
u (\eta, B_{\phi}/B_{\perp} ) = \eta
\int_{1}^{\infty} \!\! d y \frac{\ln y}{y^{2 B_{\phi}/B_{\perp}}}
\left[ v(y,\eta) - \frac{2}{y^2 - 1} \right],
\end{equation}
with
\begin{eqnarray}
v( y, \eta ) & = & \frac{2}{ ( y^2 - 1 ) } 
\frac{ [ (1 + 2 \eta + 2 \eta^2)(y^2 - 1) + 4 \eta 
]}{[ (1 + 2 \eta)^2 (y^2 - 1) + 8 \eta ]}
\nonumber \\
 & - & \left( \frac{\eta}{ f^{3/2} } [ (1 + 2 \eta)
(y^2 - 1)^2 + (8 + 4 \eta)(y^2 - 1) + 8 ] \right.
\nonumber \\ 
 & \times & \ln \left. \left\{ \frac{ [ \sqrt{f} - 2 \eta (y^2 - 1) ]^2 -
(y^2 - 1)^2 }{ [ \sqrt{f} + 2 \eta (y^2 - 1) ]^2 - (y^2 - 1)^2 }
\right\} \right);
\nonumber \\
f( y, \eta ) & = & (y^2 - 1) [ (1 + 2 \eta)^2 (y^2 - 1) + 8 \eta ].
\end{eqnarray}

Eqs. (76)-(78) are the main results of this section. They give the shift
in conductance from the standard weak-localization lineshape
due to the applied in-plane field $B_{\parallel}$. {\it Remarkably
$ \delta g $ is independent of $l_{e}$.}

Since the expression for $ \delta g $ is complicated, it is
instructive to examine several special cases. First, by
taking the limit $B_{\perp} \rightarrow 0$ it is possible
to recover the result of the previous subsection, Eq. (67).
Next consider the strong perpendicular field limit $B_{\perp}
l_{\phi}^{2} \gg \phi_{{\rm sc}}$. In this limit
\begin{equation}
\delta g = -2 \frac{e^2}{\hbar} \left( 
\frac{B_{\parallel} \Delta L}{ \phi_{{\rm sc}} } \right)
h(\eta)
\end{equation}
with
\begin{equation}
h(\eta) = \eta \int_{1}^{\infty} \!\! d y \ln y \; v(y,\eta) 
- \frac{\pi^2}{4} \eta.
\end{equation} 
A plot of $h(\eta)$ is shown in Fig. 5. It has the
asymptotic behavior
\begin{eqnarray}
h(\eta) & \approx & \eta^2 \hspace{3mm} {\rm for} \hspace{3mm}
\eta \ll 1 \nonumber \\
 & \approx & - \frac{\pi^2}{8} \hspace{2mm} {\rm for} \hspace{3mm}
\eta \gg 1.
\end{eqnarray}
Note that $h(\eta)$ changes sign; it vanishes for
$\eta \approx 0.558$.

\begin{figure}[t]
\begin{center}
\leavevmode
\epsfxsize=3.0in \epsfbox{heta.epsi}
\end{center}
\caption{Plot of $h(\eta)/\eta$ as a function of 
$\eta = \phi_{{\rm sc}}/(2 \pi L^2 B_{\perp})$. The
inset shows the small $\eta$ behavior and the zero of $h(\eta)$
in detail. $h(\eta)$ describes the perpendicular field dependence
of the in-plane magnetoresistance in the geometric limit, $l_e \ll
L \ll l_{\phi}$; see Eq. (79). }
\end{figure}

Eq. (79) reveals that for $ B_{\perp} l_{\phi}^2 \gg \phi_{{\rm sc}}$,
$\delta g$ depends only on the magnetic fields $B_{\parallel}$ and
$B_{\perp}$ and geometric parameters that characterize the
rough interface, $\Delta$ and $L$. { \it It is independent not only
of $l_e$ but also of $l_{\phi}$}. The intermediate regime in
which $l_e \ll L $, but $L \ll l_{\phi}$, is particularly
interesting. In this limit $B_{\phi} \ll \phi_{{\rm sc}}/L^2$
and hence Eq. (79) applies in a range of $B_{\perp}$ over
which $\eta$ varies from large to small. Thus the perpendicular
field magnetoconductance curves at different values of $B_{\parallel}$
cross at a point determined by the purely geometric condition
\begin{equation}
B_{\perp} L^2 \approx 1.79 \frac{h}{e}.
\end{equation}
For this reason we call this intermediate regime, $ l_e \ll L \ll
l_{\phi}$, the {\it geometric} regime.

The opposite limit of extremely long-ranged roughness, $ L \gg l_{\phi}
\gg l_e $, we call the inhomogeneous broadening regime. In this
limit, too, perpendicular magneto-conductance curves corresponding to
different small $B_{\parallel}$ cross, but the crossing point is not
determined by the purely geometric condition, Eq. (82). In this case
the range of $B_{\perp}$ over which the simplified formula, Eq. (79),
applies corresponds to very small $\eta$ and does not include the
geometric crossing point, $ \eta = 0.558 \dots$ We must therefore
use the general expression Eq. (76) to determine the crossing point.
Nonetheless in the inhomogeneous broadening regime the deviation
from the conventional weak-localization lineshape and the existence
of a crossing point has a simple interpretation. Since the random
magnetic field is correlated over a length scale $L$, the sample
breaks up into blocks of size $l_{\phi}$ each of which sees a
slightly different (but uniform) perpendicular magnetic field,
$B_{\perp} + \delta B$. For small $ \delta B$, the shift in 
conductance of a particular block may be computed from the
conventional weak-localization lineshape $g_{{\rm WL}} ( B_{\perp} )$
by expanding in a Taylor series:
\begin{equation}
\delta g \approx \frac{ \partial }{ \partial B_{\perp} } g_{{\rm WL}}
(B_{\perp}) \delta B + \frac{1}{2} \frac{ \partial^2 }{ \partial
B_{\perp}^{2} } g_{{\rm WL}} (B_{\perp}) \delta B^2 + \ldots
\end{equation}
Bearing in mind that the average value of the random field is
zero and the typical value, $B_{\parallel} \Delta / L$, the
average deviation is
\begin{equation}
\delta g \sim B_{\parallel}^2 \frac{ \Delta^2 }{ L^2 } 
\frac{ \partial^2 }{ \partial B_{\perp}^2 } g_{{\rm WL}} 
( B_{\perp} ).
\end{equation}
The curvature of $g_{{\rm WL}}(B_{\perp})$ is positive 
at $B_{\perp} = 0$. At large $B_{\perp} \gg \phi_{{\rm sc}}/
l_{\phi}^2$, $g_{{\rm WL}} (B_{\perp}) \sim (e^2/h) \ln
(B_{\perp}/B_{\phi})$; hence the curvature is negative.
Thus there must be an inflection point at which the curvature
of $g_{{\rm WL}}(B_{\perp})$ vanishes. According to Eq. (84) 
this is the crossing point. Substitution of the known asymptotic
forms of $g_{{\rm WL}} (B_{\perp})$ in Eq. (84) allows a check
on the consistency of this interpretation. For small $B_{\perp}$
we recover the estimate given in the second line of Eq. (70).
For $B_{\perp} \gg \phi_{{\rm sc}}/l_{\phi}^2$ we obtain
\begin{equation}
\delta g \sim \frac{e^2}{h} \frac{ B_{\parallel}^2 \Delta^2 }{ 
B_{\perp}^2 L^2 }
\end{equation}
in agreement with the small $ \eta $ limit of Eq. (79).

\section{Comparison to Experiment and Related Work}

In their experiments Wheeler and collaborators apply a
fixed $B_{\parallel}$ and measure the conductance as a
function of $B_{\perp}$. They have reported such lineshape
measurements for many devices and have also explored
the effect of varying electron density on the lineshape.
The data of Anderson {\em et al.} \cite{bob2} are particularly
interesting because they have independently measured the 
interface roughness by etching away the oxide 
and imaged the exposed silicon surface using atomic force
microscopy. Hence we focus on the measurements reported in that
paper.

Fig. 6 shows the measured magnetoconductance of the control device
at $B_{\parallel}=0$ (grey) and $B_{\parallel}=1.5$ T (black).
By fitting the $B_{\parallel} = 0$ data to a weak-localization lineshape
(dashed curve in Fig. 6), Eq. (44), Anderson {\em et al.} infer $l_e=$ 0.085 
$\mu{\rm m}$, $l_{\phi}= 0.76$  $\mu{\rm m}$. Atomic force microscope
measurements yield $\Delta = 1.35 {\rm \AA}$ and
$L= 0.055$ $\mu{\rm m}$. Thus the device is in the short-range correlated
regime. The black curve shows the calculated lineshape using
the independently measured values of $\Delta$, $L$, $l_e$
and $l_{\phi}$. No parameters are adjusted except for a 
constant offset. The fit is good, comparable, for example,
to the fit to weak-localization lineshape at $B_{\parallel}=0$.

\begin{figure}[t]
\begin{center}
\leavevmode
\epsfxsize=3.0in \epsfbox{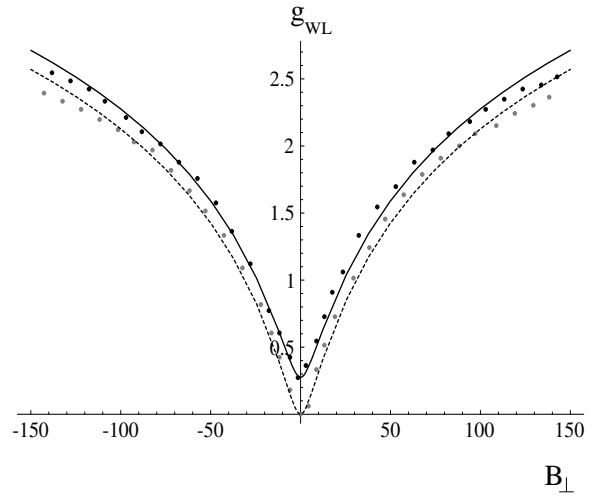}
\end{center}
\protect\caption{Plot of the conductance (in units of $10^{-5}$ mhos)
against perpendicular field, $B_{\perp}$ (in Gauss), for the control
device of Anderson {\em et al.} 
\protect\cite{bob2}. 
Data points for $B_{\parallel}
=0$ are grey; for $B_{\parallel} = 1.5$ T, black. The dashed line is
a fit to the weak-localization lineshape, from which Anderson {\em et al.}
infer $l_e = 0.085 \mu{\rm m}$, $l_{\phi} = 0.76 \mu{\rm m}$. The
solid curve shows the lineshape at $B_{\parallel} = 1.5$ T calculated
from the expression for homogeneous broadening derived in this paper,
Eqs. (44) and (55). All the parameters for this curve are independently
determined. The values of $\Delta = 1.35 {\rm \AA}$ and $L = 0.055 \mu{\rm m}$
are obtained from atomic force microscope images; for $l_e$ and $l_{\phi}$,
we use the values cited above.}
\end{figure}

Anderson {\em et al.} also report measurements on a second
device (called the textured device). This device is fabricated
on a silicon wafer that has been roughened by argon sputtering
prior to gate oxidation. As a result the textured device has
atypical values of $\Delta$ and $L$. $L$ for the textured sample
is comparable to $l_e$. Consistent with this we do not obtain
a good zero-parameter fit to either the short or long-range 
roughness expressions derived here. Following Anderson {\em et al.}
we can force-fit the data to the homogeneously broadened lineshape;
but the value of $\Delta^2 L$ we then obtain is inconsistent with
the atomic force microscope images.

Finally it is interesting to compare the experiments of Wheeler 
{\em et al.} to other work on random or periodic magnetic fields.
Such fields have been realized by gating a GaAs/AlGaAs heterostructure
with type II superconducting film \cite{bending,geim}; 
by randomly depositing type I
superconducting grains \cite{smith} or ferromagnetic dysprosium dots
\cite{weiss} on the
heterostructure surface; and by covering the device with a rough
macroscopic magnet \cite{mancoff}. In work most closely related to Wheeler 
and co-workers, Gusev {\em et al.} have prepatterned the substrate
by drilling a lattice of holes so that the interface at which the
two-dimensional electron gas forms is periodically modulated \cite{gusev}.
An inhomogeneous magnetic field leads to many interesting 
features in the magnetoresistance even in the classical and
ballistic regimes; many experiments have focused on these
effects. Except for the work of Rammer and Shelankov \cite{theory} 
noted below, much of the theoretical literature too explores these
regimes or focuses on the fundamental question of whether all
eigenstates are localized in a random magnetic field.

Two experiments study quantum corrections in the
weak-localization limit: In the work of Bending, von Klitzing
and Ploog \cite{bending} the inhomogeneous magnetic field is produced by an
Abrikosov lattice of flux lines in the superconducting gate;
hence it is periodic. However the magnetoresistance is not
sensitive to the arrangement of the vortices since the dephasing
length is shorter than the distances between the vortices.
This circumstance was analyzed by Rammer and Shelankov \cite{theory}.
In the experiment of Gusev {\em et al.} too the magnetic field
is essentially periodic rather than random allowing them to 
analyze it via a chessboard model \cite{gusev}. Thus in both these works
the inhomogeneous magnetic field is rather different from
the random Gaussian correlated field studied in the experiments of
Wheeler and co-workers and analyzed here.

\section{Summary and Conclusion}  

In this paper we have been concerned with the influence of
interface roughness on the magnetoconductance of MOSFETs. 
Using a Born-Oppenheimer approximation we have shown that 
effectively electrons respond to an applied in-plane magnetic
field as though it were a random perpendicular field. Technically,
the random magnetic field appears as a Berry phase term in the
Born-Oppenheimer effective Hamiltonian. Eq. (9) shows that the
random magnetic field is the component of $B_{\parallel}$ normal
to the rough surface ${\cal Z}_{g}(x,y)$. The Born-Oppenheimer analysis
also gives precise meaning to the surface ${\cal Z}_{g}(x,y)$: it is
the location of the centre of the local subband wavefunction. 

The next step is to analyze the magnetoconductance. This is controlled
by the familiar physics of weak-localization: at zero field there is
a contribution to the conductance due to constructive interference 
of closed electron paths and their time-reversed counterparts. 
This contribution is suppressed by application of a perpendicular
magnetic field because the electron acquires an Aharonov-Bohm phase
as it traverses a closed path. The analysis of this paper shows that
an in-plane field too has an effect due to interface roughness.
The electron acquires a Berry's phase equal to the flux of the
random field through the closed path. The effect of the random 
Berry's phase on the conductance is the main focus of this paper. 

To calculate the effect it is necessary to sum the standard
divergent series of Cooperon diagrams (Fig. 3). We 
find it useful to reformulate
this sum in a manner reminiscent of the central limit theorem.
Feynman graphs of high order control the sum. In these graphs
we are able, without significant error, 
to replace pairs of Green function lines with
suitable Gaussian factors. At zero
field, the $n^{{\rm th}}$ order diagram is then essentially the
weight of a closed $n$-step 
Gaussian random walk [Eq. (27)]. For the known
case of an applied perpendicular field, this formulation brings
out clearly the relationship between the classic weak-localization
lineshape (Eq. 44) and the distribution of directed area for 
closed random walks (see Eqs. 38 and 39 and Appendix C). The method also proves
useful when an in-plane field is applied for long-range correlated
roughness ($L \gg l_e$). In this case the magnetoconductance is
related to the directed area distribution of an {\em interacting}
random walk (see Eqs. 63 and 73). To make contact with experiment
it is sufficient to analyze this problem perturbatively. We do
not attempt a complete solution of the interacting random walk
or calculation of the full magnetoconductance for long-range
correlated roughness. These problems are left open for future
work.

For short-range correlated roughness, however, we are able to
obtain the full magnetoconductance. The dependence of the
conductance on perpendicular field, $B_{\perp}$, still has
the classic weak-localization lineshape (Eq. 44) when an in-plane
field is applied, just as it does at $B_{\parallel} = 0$. The
only effect of the in-plane field is to enhance the dephasing
rate (Eq. 52). In terms of field scales, $B_{\phi}$ becomes 
larger; the effective $B_{\phi}$ is given by Eq. (55). Hence
we say that short-range correlated roughness produces homogeneous
broadening of the weak-localization lineshape.

The departure from the conventional lineshape is more striking
in case of long-range correlated roughness (although quantitatively
smaller if all parameters except $L$ are held fixed). The change
in conductance, $\delta g$, when a parallel field is applied
(``parallel field magnetoconductance'') is given by Eq. (76).
The most remarkable feature of this expression is that it is
independent of $l_e$. $\delta g$ is determined entirely by the
magnetic field, geometric parameters of the interface, $\Delta$
and $L$, and the dephasing length, $ l_{\phi}$. It may be of
either sign (positive near $B_{\perp} = 0$ and negative for
large $B_{\perp}$). The expression for $\delta g$ is lengthy
and cannot be more compactly expressed in terms of known
special functions. It can be numerically computed and plotted
with ease and precision; however it is worthwhile to examine
several special cases.

In the limit of extremely long-ranged roughness, $L \gg l_{\phi} 
\gg l_e$, the parallel field magnetoconductance has a simple
interpretation: the sample breaks up into independent blocks
of size $l_{\phi}^2$ each of which sees a slightly different
magnetic field. Hence in this limit we say that the effect of
an in-plane field and roughness is to inhomogeneously broaden
the weak-localization line. Pursuing this interpretation we can
understand the sign change in $\delta g$ and recover
its form in various circumstances [see the discussion following
Eqs. (70) and (82)]. 

The limit of intermediate-range correlated roughness, $l_{\phi}
\gg L \gg l_e$, is particularly striking. For sufficiently
large perpendicular field, $B_{\perp} l_{\phi}^2 \gg \phi_{{\rm sc}}$,
$\delta g$ becomes independent not only of $l_e$ but also $l_{\phi}$
[see Eq. (79)]. Essentially the parallel magnetoconductance is
controlled by the function $h(\eta)$ where $ \eta = \phi_{{\rm sc}}/
(2 \pi L^2 B_{\perp})$ is a measure of the perpendicular flux (see
Fig. 5). 

Another special case, useful for making contact with experiment
\cite{albert} is the parallel field magnetoconductance for $B_{\perp}=0$
(Eq. 67). In this case, $\delta g$, in units of $e^2/h$, is 
a product of the squared flux
$(B_{\parallel} \Delta L)^2$, determined entirely by the field
and the geometry of the interface, and the function $f(\alpha)$.
$f(\alpha)$ describes the crossover as $l_{\phi}$ is varied
relative to $L$ ($\alpha = l_{\phi}^2/L^2$. See Fig. 4). 

Wheeler and co-workers have shown experimentally 
in a number of cases that the
effect of an in-plane field is to produce homogeneous broadening.
Our analysis of the short-range correlated regime provides a 
theoretical justification of such fits to data and allows quantitative
information about the interface roughness to be extracted. It is
encouraging that for the control sample studied by Anderson {\em 
et al.} \cite{bob2}, for which the interface parameters are 
independently measured via atomic force microscopy, 
our formulae give a satisfactory
zero parameter fit (Fig. 6). Unfortunately, our analysis shows that
in the short-range regime, magnetoresistance measurements will not
separately yield $\Delta$ and $L$: instead they provide the 
combination $\Delta^2 L$. Thus it will be necessary to 
combine magnetoconductance measurements with other techniques to 
separately obtain $\Delta$ and $L$.

The roughness of the interface is controlled by its processing.
For example, Anderson {\em et al.} \cite{bob2} are able to increase
both $\Delta$ and $L$ by argon sputtering the silicon surface
prior to gate oxidation. It would be desirable to create devices
with long-range correlated roughness in order to experimentally
observe, for example, the geometric regime described above. At 
the same time, for long-range correlated roughness, the prospects
are much better for extracting both $\Delta$ and $L$ from 
magnetoconductance measurements alone, provided the measurements
can be made with sufficient precision. Eq. (76) provides the
means to study the feasibility of such experiments.

Another potential application of our analysis is related to
efforts to engineer the MOSFET interface to have specific
structure, for example, a structure that is periodic in
one direction like a corrugated sheet \cite{sakaki}. Weak-localization
magnetoconductance measurements would provide a non-destructive
way to check that structures have been successfully fabricated.
It should be a straightforward and interesting extension of our
analysis to calculate the magnetoresistance signature of various
simple periodic structures.

\section*{Acknowledgements}

It is a pleasure to acknowledge many stimulating conversations
with Albert Chang, Rachel Lombardi, Satya Majumdar,
Don Monroe, and Bob Wheeler. H. Mathur
is supported by NSF Grant DMR 98-04983 and an Alfred P Sloan
Research fellowship and acknowledges the hospitality of the 
Aspen Center for Physics.

\appendix

\section{Adiabatic Approximation}

Here we briefly discuss the conditions for the
validity of the Born-Oppenheimer approximation used
in Section II. Roughly, we want only the low-lying states
of the lowest subband to be occupied and we want the gap
between subbands to be large. For this purpose, and to ensure
the two dimensionality of the electron gas, we need the Fermi
energy to be small compared to the subband spacing. The subband
spacing is of order $\hbar^2/m \ell^2$ where $\ell$ is the length
scale over which the electron is confined. Hence this condition
may be phrased as
\begin{equation}
\lambda_{f} \gg \ell.
\end{equation}
To obtain additional conditions we return to the Schr\"{o}dinger
Eq. (1) and seek a solution of the form
\begin{equation}
\Psi (x,y,z) = \sum_{n} \psi_{n} (x,y) \phi_{n}(z;x,y).
\end{equation}
For simplicity we limit discussion to the case of zero magnetic 
field. $\psi_{n} (x,y)$ then obeys \cite{geomphase,shankar}
\begin{eqnarray}
& & \left[ - \frac{\hbar^2}{2m} \nabla^2 + U_{{\rm imp}}(x,y)
+E_{n}(x,y) + W_{n}(x,y) \right] \psi_{n}(x,y) 
\nonumber \\
& & + \sum_{m \neq n} H_{nm} \psi_m (x,y) = E \psi_n (x,y). 
\nonumber \\
\end{eqnarray}
Here
\begin{equation}
H_{nm} = \frac{\hbar^2}{m} \langle n; x,y | \nabla | m; x,y \rangle. 
\nabla + \frac{\hbar^2}{2m} \langle n; x,y | \nabla^2 | n; x,y \rangle.
\end{equation}
Eq. (A4) constitutes an exact reformulation of the Schr\"{o}dinger
Eq. (1). The Born-Oppenheimer approximation consists of keeping only
one term in the sum (A2) and hence omitting the off-diagonal terms
in (A3). 

In a commonly used model of the interface \cite{ando} it is assumed
that the confining potential has the form
\begin{equation}
U_{{\rm conf}} (z; x,y) = {\cal U}[ z - \zeta(x,y) ].
\end{equation}
Here $\zeta(x,y)$ is the elevation of the silicon-oxide boundary.  In this
model the mean elevation of the local subband wavefunction exactly tracks
the interface; hence the surface ${\cal Z}_g (x,y)$, probed by
weak-localization magnetoresistance, has the same statistics as the
silicon-oxide interface $\zeta(x,y)$, which is presumably imaged by atomic
force microscopy. For a more general form of the confinement potential
this precise correspondence between ${\cal Z}_{g} (x,y)$ and $\zeta (x,y)$
is lost but it seems reasonable that the two surfaces would have similar
statistics, comparable correlation lengths and mean-square fluctuations.

Eq. (A5) implies that the local subband wave function is of the form 
\begin{equation}
\phi_{n}(z;x,y) = \xi_n [ z - \zeta(x,y) ].
\end{equation}
Using (A6) we can estimate the off-diagonal elements given
by (A4). Requiring these to be small compared to subband
spacing $\hbar^2/m \ell^2$, leads to 
\begin{equation}
\ell \nabla^2 {\cal Z}_{g} \ll 1, 
\hspace{4mm}
\nabla {\cal Z}_g \ll 1.
\end{equation}
In summary, Eqs. (A1) and (A7) are the conditions for the
validity of the Born-Oppenheimer approximation. Note that
the same conditions can in fact be derived under less
restrictive assumptions than Eq. (A5).

If we assume that ${\cal Z}_{g}$ is statistically the same as the surface 
imaged by atomic force microscopy, Eq. (A7) may be rewritten 
\begin{equation}
\frac{ \ell \Delta_{{\rm AFM}} }{ L^2 } \ll 1,
\hspace{4mm}
\frac{ \Delta_{{\rm AFM}} }{L} \ll 1.
\end{equation}
For the control sample studied by Anderson {\em et al.} \cite{bob2}
$\Delta_{{\rm AFM}} = 1.35~{\rm \AA}$, $L = 0.055~\mu{\rm m}$, 
$\lambda_{f} = 100~{\rm \AA}$; for the textured
sample, $\Delta_{{\rm AFM}} = 7.5~{\rm \AA}$, $L = 0.09~\mu{\rm m}$, 
$\lambda_f = 70~{\rm \AA}$. Taking $\ell \approx 30~{\rm \AA}$, it is 
easily seen that the Born-Oppenheimer approximation is applicable.

\section{Cooperon Path Integral}

The purpose of this Appendix is to calculate the integral,
Eq. (36). It is useful to first calculate
\begin{eqnarray}
D^{(n)}(y,y') & = & \int_{-\infty}^{+\infty} d y_1 \ldots
d y_n D^{(0)} (y,y_1) \ldots D^{(0)} (y_n, y');
\nonumber \\
D^{(0)} (y,y') & = & \frac{1}{\sqrt{\pi}} e^{ - (y - y' )^2 - g
(y + y')^2 }.
\end{eqnarray}
It is easy to show by induction that $D^{(n)}$ is of the form
\begin{equation}
D^{(n)}(y,y') = \frac{1}{\sqrt{\pi}} \frac{1}{A_{n}} 
e^{- a_n (y-y')^2 - b_{n} (y+y')^2 }
\end{equation}
with $ a_n b_n = g$. By completing squares we find
\begin{eqnarray}
D^{(n+1)}(y,y') & = & \int_{-\infty}^{+ \infty} d y_1 D^{(0)}
(y, y_1) D^{(n)}(y_1, y') \nonumber \\
 & = & \frac{1}{\sqrt{\pi}} \frac{1}{A_{n+1}} e^{ - a_{n+1} (y-y')^2
- b_{n+1} (y+y')^2 }
\end{eqnarray}
provided $a_n b_n = g$. Here
\begin{eqnarray}
\frac{1}{a_{n+1}} & = & \frac{1}{a_n + g} + 
\frac{1}{b_n + 1}; \nonumber \\
\frac{1}{b_{n+1}} & = & \frac{1}{a_n + 1} +
\frac{1}{b_n + g} ; \nonumber \\
A_{n+1} & = & A_{n} \sqrt{ a_n + b_n + 1 + g }.
\end{eqnarray}
Using Eq. (B4) we can verify that $a_{n+1} b_{n+1} = g$
ensuring $D^{(n+2)}$ will also be of the form given in
Eq. (B2). In principle the recurrence relations, Eq. (B4), 
determine $a_n$, $b_n$ and $A_n$; but in practice it is 
easier to follow a different method. 

Experience with paths integrals suggests that the 
eigenfunctions of the kernel $D^{(0)}$ are $H_{m}( \sqrt{2 \alpha} y )
e^{- \alpha y^2}$ with $\alpha$ suitably adjusted---of the same
form as the eigenfunctions of a harmonic oscillator.
To verify this conjecture and to fix $\alpha$ we use a 
generating function for the Hermite polynomials
\begin{equation}
\exp \left( - x^2 + 2 x \sqrt{2 \alpha} y - \alpha y^2 \right)
= \sum_{m=0}^{\infty} \frac{x^m}{m!} H_{m} ( \sqrt{2 \alpha} y )
e^{ - \alpha y^2 }.
\end{equation}
Upon completing squares we find
\begin{eqnarray}
& & 
\int d y' D^{(0)} (y,y') \exp ( - x^2 + 2 x \sqrt{2 \alpha} y' 
- \alpha y'^2 ) \nonumber \\
& &
= \frac{1}{1 + \sqrt{g}} e^{\left[ 
- \left( \frac{1 - \sqrt{g} }{1 
+ \sqrt{g}} \right)^2 x^2 + 2 \left( \frac{1 - \sqrt{g} }{1 +
\sqrt{g}} \right) x 2 g^{1/4} y - 2 \sqrt{g} y^2 
\right]} \nonumber \\
& &
= \sum_{m=0}^{\infty} \frac{ ( 1 - \sqrt{g} )^m }{ ( 1 + \sqrt{g} 
)^{m+1} } \frac{x^m}{m!} H_{m} ( 2 g^{1/4} y ) \exp( - 2 \sqrt{g} y^2 )
\end{eqnarray}
provided $\alpha = 2 \sqrt{g}$. This condition is imposed by 
requiring that the coefficient of the $y^2$ term in the second
line of Eq. (B6) should be $ \alpha $. Comparing Eqs. (B5) and  
(B6) we conclude that $H_{m} ( 2 g^{1/4} y ) e^{- 2 \sqrt{g} y^2 }$
are eigenfunctions of $D^{(0)}$ with eigenvalue 
$ (1 - \sqrt{g})^m/(1 + \sqrt{g} )^{m+1}$ for $ m = 0,1,2, \ldots$ 

Now let us evaluate
\begin{eqnarray}
& &
\int_{-\infty}^{+\infty} d y' D^{(n)}(y,y') \exp( -2 \sqrt{g} y'^2
- x^2 + 4 g^{1/4} x y ) \nonumber \\
& & =
\frac{1}{A_n} \frac{ \sqrt{ \pi a_n } }{a_n + \sqrt{g}} e^{\left[
- 2 \sqrt{g} y^2 - \left( \frac{a_n - \sqrt{g}}{a_n + \sqrt{g}} \right)^2
x^2 + 4 g^{1/4} \left( \frac{a_n - \sqrt{g}}{a_n + \sqrt{g}} \right) x y
\right]} \nonumber \\
& & = 
\sum_{m=0}^{\infty} \frac{1}{m!} \frac{ \sqrt{ \pi a_n } }{A_n}
\frac{ ( a_n - \sqrt{g} )^m }{ ( a_n + \sqrt{g} )^{m+1} }
x^m H_{m} (2 g^{1/4} y) e^{-2 \sqrt{g} y^2} \nonumber \\
& & =
\sum_{m=0}^{\infty} \frac{1}{m!} \frac{ (1 - \sqrt{g})^{m(n+1)}}{
(1 + \sqrt{g})^{(m+1)(n+1)} } x^m H_{m}(2 g^{1/4} y) e^{-2 \sqrt{g} y^2}.
\end{eqnarray}
The integral in the first line of Eq. (B7) has been analyzed in two ways.
First we use the ansatz for $D^{(n)}$, Eq. (B2), perform the integral by
completing squares, and expand the result using the generating formula
for the Hermite polynomials, Eq. (B5). The results are given in the
second and third lines. Alternatively, we expand the exponential in the
first line in terms of the eigenfunctions of $D^{(0)}$ using the generating
formula, Eq. (B5), and note that the eigenfunctions of $D^{(0)}$ are also
eigenfunctions of $D^{(n)}$ but with the eigenvalue raised to the $(n+1)^{{\rm
st}}$ power (considered as an integral operator, $D^{(n)}$ amounts to
$n+1$ repeated applications of $D^{(0)}$). The result is given in the
fourth line. Comparison of the third and fourth lines reveals
\begin{eqnarray}
a_n & = & \sqrt{g} \left[ \frac{ (1 + \sqrt{g})^{n+1} + (1 - \sqrt{g})^{n+1} 
}{ (1 + \sqrt{g})^{n+1} - (1 - \sqrt{g})^{n+1} } \right] \nonumber \\
b_n & = & \sqrt{g} \left[ \frac{ (1 + \sqrt{g})^{n+1} - (1 - \sqrt{g})^{n+1}
}{ (1 + \sqrt{g})^{n+1} + (1 - \sqrt{g})^{n+1} } \right] \nonumber \\
A_{n} & = & \frac{ \sqrt{\pi} }{2 g^{1/4}} \left[ (1 + \sqrt{g})^{n+1} +
(1 - \sqrt{g})^{n+1} \right]^{1/2} \nonumber \\
 & & \times \left[ (1 + \sqrt{g})^{n+1} - (1 - \sqrt{g})^{n+1} \right]^{1/2}.
\end{eqnarray}
These expressions are the solution to the recurrence relation in Eq. (B4)
with the initial condition $a_0 =1, b_0=g, A_0=1$. 

In summary the integral $D^{(n)}$, defined in Eq. (B1) is given by
Eqs. (B2) and (B8).

Now we evaluate Eq. (36) for $K_{B_{\perp}}^{(n)}({\mathbf r},{\mathbf r}')$.
Rescaling the co-ordinates we obtain
\begin{equation}
K_{B_{\perp}}^{(n)}({\mathbf r},{\mathbf r}') = \frac{1}{2 l_{e}^{2}} E^{(n)}
({\mathbf r},{\mathbf r}')
\end{equation}
with
\begin{eqnarray}
E^{(n)}({\mathbf r},{\mathbf r}') & = & \int d {\mathbf r}_1 \ldots d {\mathbf r}_n
E^{(0)}({\mathbf r},{\mathbf r}_1) \ldots E^{(0)}({\mathbf r}_n,{\mathbf r}'),
\nonumber \\
E^{(0)}({\mathbf r},{\mathbf r}') & = & \frac{1}{\pi} \exp \left[ - |{\mathbf r} -
{\mathbf r}'|^2
- i \beta (x-x')(y+y') \right]
\end{eqnarray}
and $\beta = B_{\perp}/B_e$. In the Landau gauge $E^{(0)}$ depends 
only on $(x-x')$ and has the Fourier transform
\begin{eqnarray}
 & &
\frac{1}{\sqrt{\pi}} \exp \left[ - (x-x')^2 - i \beta (y+y')(x-x') \right]
\nonumber \\
& & 
= \int_{-\infty}^{+\infty} \frac{ d k }{2 \pi} \exp \left( - 
\frac{ [ k + \beta (y+y') ]^2 }{4} \right) e^{i k (x-x')}.
\end{eqnarray}
By virtue of translational invariance
\begin{eqnarray}
E^{(n)}({\mathbf r},{\mathbf r}') & = & \int_{-\infty}^{+\infty} \frac{dk}{2 \pi}
\int_{-\infty}^{+\infty} d y_1 \ldots d y_n \frac{1}{\sqrt{\pi}^{n+1}}
e^{i k (x-x')} \nonumber\\
& &
\times
e^{-[k+\beta(y+y_1)]^2/4} \ldots e^{-[k+\beta(y_n+y')]^2/4} \nonumber \\
& & 
\times e^{-(y-y_1)^2} \ldots e^{-(y_n - y')^2}.
\end{eqnarray}
If we shift $y_i \rightarrow y_i - k/2\beta$
\begin{eqnarray}
E^{(n)}({\mathbf r},{\mathbf r}') & = & \int_{-\infty}^{+\infty} \frac{dk}{2 \pi}
\int_{-\infty}^{+\infty} d y_1 \ldots d y_n \frac{1}{\sqrt{\pi}^{n+1}}
e^{i k (x-x')} \nonumber\\
& &
\times \exp \left[ - \frac{\beta^2}{4} ( \overline{y} + y_1 )^2 +
\ldots - \frac{\beta^2}{4} (y_n + \overline{y}' )^2 \right]
\nonumber \\
& &
\times e^{-(\overline{y}-y_1)^2 + \ldots - (y_n - \overline{y}')^2 }
\end{eqnarray}
Here $ \overline{y} = y - k/2\beta$, $ \overline{y}' = y' - k/2\beta$
and the $k$-integral must be performed after the $y$-integrals. Evidently
\begin{equation}
E^{(n)} ({\mathbf r},{\mathbf r}') = \int_{-\infty}^{+\infty} \frac{dk}{2\pi}
e^{i k (x-x')} D^{(n)} \left( y - \frac{k}{2 \beta}, y' - 
\frac{k}{2 \beta} \right).
\end{equation}
$D^{(n)}$ is given by Eqs. (B2) and (B8) with $ g \rightarrow \beta^2/4$.
Performing the remaining Gaussian integral over $k$ yields
\begin{eqnarray}
E^{(n)} ({\mathbf r},{\mathbf r}') & = & \frac{1}{\pi q_{n}} 
e^{- a_n |{\mathbf r}-{\mathbf r}'|^2 - i \beta (y+y')(x - x') }; \nonumber \\
a_n & = & \frac{ [ 1 + (\beta/2) ]^{n+1} + [ 1 - (\beta/2) ]^{n+1} 
}{ [ 1 + (\beta/2) ]^{n+1} - [ 1 - (\beta/2) ]^{n+1} }; \nonumber \\
q_n & = & \frac{1}{\beta} \left[ \left(1 + \frac{\beta}{2} \right)^{n+1}
- \left( 1 - \frac{\beta}{2} \right)^{n+1} \right].
\end{eqnarray}
Eqs. (B9), (B10) 
and (B15) constitute the final expression for $K_{B_{\perp}}^{(n)}
({\mathbf r},{\mathbf r}')$.

The calculation of $K_{B_{\perp}}^{(n)}$ is similar to the solution
of Schr\"{o}dinger's equation for an electron in a magnetic field 
because the slowly varying eigenfunctions of $K_{B_{\perp}}^{(0)}$
obey the differential Eq. (33). However the derivation of Eq. (B15)
goes beyond solution of the differential equation because we also
obtain the short distance behavior of $K_{B_{\perp}}^{(n)}$.

Setting $ {\mathbf r} = {\mathbf r}'$ in Eq. (B15) we obtain 
\begin{eqnarray}
K_{B_{\perp}}^{(n-1)} ({\mathbf r},{\mathbf r}') & = & \frac{1}{2 \pi l_e^2}
\frac{B_{\perp}}{B_e} \left[ \left(1 + \frac{B_{\perp}}{2 B_e} \right)^n
- \left( 1 - \frac{ B_{\perp} }{ 2 B_e } \right)^n \right]^{-1} 
\nonumber \\
 & \approx & \frac{B_{\perp}}{2 \phi_{{\rm sc}}} 
\left[ \sinh \left( \frac{ n B_{\perp}}{2 B_e} \right) \right]^{-1}
\end{eqnarray}
provided that $B_{\perp}/B_e$ is sufficiently small that not only
is $B_{\perp}/ B_e \ll 1$ but also $ (B_{\perp}/B_e)^2 n \ll 1$.
The approximate expression in Eq. (B16) is the same as Eq. (41). For
a given large $B_e/B_{\perp}$ Eq. (41) is accurate only for $n \ll 
(B_e/B_{\perp})^2$; but since $K_{B_{\perp}}^{(n)}({\mathbf r},{\mathbf r}')$
is exponentially small for $ n \gg B_e/B_{\perp}$, no significant
error is made by taking Eq. (41) to apply for all $n$.

\section{Directed Area Distribution}

First let us make precise the meaning of directed area. Imagine
an $n$-sided polygon drawn on a plane. The polygon is endowed with
an orientation by assigning each edge a direction such that each
vertex has one incoming edge and one outgoing. Roughly, the two
orientations of a given polygon correspond to the two senses in
which its perimeter may be circumnavigated. 

Now suppose a magnetic field is applied perpendicular to the
plane. Roughly, the directed area is the quantity by which the
magnetic field must be multiplied in order to obtain the flux
through the oriented polygon (in magnitude and sign). For a
simple polygon with no self-intersection the directed area is
equal in magnitude to the area; the sign is positive or negative
depending on whether the orientation of the polygon is 
anti-clockwise or clockwise viewed from the direction in
which the magnetic field points. For a polygon with 
self-intersection, the directed area is obtained by weighting
the area of each infinitesimal area element by its winding 
number before adding them. Here the winding number of an area
element is computed by drawing a vector from the element to
the boundary of the polygon and counting the number of 
anticlockwise turns made by the vector as its tip circumnavigates
the perimeter (clockwise turns count as negative turns). For a
polygon with successive vertices at ${\mathbf r},{\mathbf r}_1, \ldots,
{\mathbf r}_{n-1}, {\mathbf r}$ immersed in a magnetic field pointing in
the $z$-direction, the directed area is given by
\begin{eqnarray}
a & = & - \frac{1}{2} (x_1 - x)(y_1 + y) -  \frac{1}{2} (x_2 - x_1)(y_2
+ y_1) + \ldots 
\nonumber \\
& & 
\ldots - \frac{1}{2} (x - x_n)(y + y_n).
\end{eqnarray}

Let the probability density of drawing an $n$-sided oriented
polygon with successive vertices at ${\mathbf r},{\mathbf r}_1, \ldots,
{\mathbf r}_{n-1}, {\mathbf r}$ be $(n/\pi^{n-1}) \exp[ - | {\mathbf r} - 
{\mathbf r}_1 |^2 + \ldots + | {\mathbf r}_{n-1} - {\mathbf r} |^2 ]$. 
Eqs. (C1) and (B10) show that
\begin{equation}
E^{(n-1)} ({\mathbf r},{\mathbf r}) = \frac{1}{n \pi} \int_{-\infty}^{+\infty}
d a P(a) \exp (i 2 \beta a).
\end{equation}
Here $P(a) =$ probability distribution of the directed area
for the Gaussian polygon ensemble defined above. By inverting
the Fourier transform in Eq. (C2) and making use of Eq. (B15)
for $E^{(n-1)}({\mathbf r},{\mathbf r}')$  we obtain
\begin{eqnarray}
P(a) & = & \int_{-\infty}^{+\infty} \frac{d \alpha}{2 \pi} \left( 
\frac{n \alpha}{2} \right) \left[ \left( 1 + \frac{\alpha}{4}
\right)^n - \left( 1 - \frac{\alpha}{4} \right)^n \right]^{-1}
e^{- i \alpha a} \nonumber \\
 & \approx & \int_{-\infty}^{+\infty} \frac{d \alpha}{2 \pi} \left(
\frac{n \alpha}{4} \right) \left[ \sinh \left( \frac{ n \alpha}{4}
\right) \right]^{-1} e^{-i \alpha a} \nonumber \\
 & = & \frac{\pi}{n} {\rm sech}^{2} \left( \frac{ 2 \pi a }{n} \right).
\end{eqnarray}
The approximate form of the integrand in the second line above
is valid only for $\alpha \ll 1/\sqrt{n}$; but it may be extended
to all $\alpha$ since the integrand is already exponentially small
for $ \alpha \gg 1/n$.

Eq. (C3) is the formula for the directed area distribution.
It coincides with Eq. (41) if we take the step size of the
random walker to be $\sqrt{2} l_e$ instead of 1 as we have 
done in this Appendix. 

The distribution of directed area for Brownian
motion was analysed by L\'evy \cite{levy} and restudied more
explicitly in connection with electron transport
by Argaman {\em et al.} \cite{argaman}. Here however we 
require the area distribution
$P_n(a)$ for a random walk with a {\em finite} number of 
steps $n$. The large $n$ behaviour of $P_n(a)$ was recently
obtained by Bellissard {\em et al.} for random polygons on
a square lattice by an interesting application of
non-commutative geometry \cite{bellissard}. Their result agrees with
the large $n$ limit of eq (C3).



\end{document}